\documentclass{aa}  

\usepackage{graphicx}
\usepackage{xcolor}
\usepackage{amsmath}
\usepackage{commath}
\usepackage{txfonts}

\begin{document} 
\def\d{\displaystyle}
\def\pa{\partial}
\def\l{\left}
\def\r{\right}
\def\c{\nabla \cdot \pmb{\xi}}
\def\e{\exp\l\{i \left( n \p + k_z z - \o t \right) \r\}}
\def\z{\zeta}
\def\a{\alpha}
\def\b{\beta}
\def\D{\Delta}
\def\k{\kappa}
\def\o{\omega}
\def\p{\varphi}
\def\s{\sigma}
\def\t{\tau}
\def\ch{\chi}

   \title{Quasimodes in the cusp continuum in nonuniform magnetic flux tubes}

   \author{M. Geeraerts \inst{1}
   		  \and
          P. Vanmechelen \inst{1}
          \and
          T. Van Doorsselaere \inst{1}
          \and
          R. Soler \inst{2,3}
          }

   \institute{Centre for mathematical Plasma Astrophysics (CmPA), KU Leuven, Celestijnenlaan 200B bus 2400, B-3001 Leuven, Belgium\\
   \and
   Departament de F\'isica, Universitat de les Illes Balears, E-07122 Palma de Mallorca, Spain\\
   \and
   Institute of Applied Computing \& Community Code (IAC3), Universitat de les Illes Balears, E-07122 Palma de Mallorca, Spain
             }

   \date{}

 
  \abstract
   {The study of magnetohydrodynamic (MHD) waves is important both for understanding heating in the solar atmosphere (and in particular the corona) and for solar atmospheric seismology. The analytical investigation of wave mode properties in a cylinder is of particular interest in this domain, as many atmospheric structures can be modelled as such in a first approximation.}
   {The aim of this paper is to use linearized ideal MHD to study quasimodes (global modes that are damped through resonant absorption) with a frequency in the cusp continuum, in a straight cylinder with a circular base and an inhomogeneous layer at its boundary which separates two homogeneous plasma regions inside and outside. We are in particular interested in the damping of these modes, and shall hence try to determine their frequency as a function of background parameters.}
   {After linearizing the ideal MHD equations, we find solutions to the second-order differential equation for the perturbed total pressure in the inhomogeneous layer in the form of Frobenius series around the regular singular points that are the Alfv\'en and cusp resonant positions, as well as power series around regular points. By connecting these solutions appropriately through the inhomogeneous layer and with the solutions of the homogeneous regions inside and outside the cylinder, we derive a dispersion relation for the frequency of the eigenmodes of the system. }
   {From the dispersion relation, it is also possible to find the frequency of quasimodes even though they are not eigenmodes. As an example, we find the frequency of the slow surface sausage quasimode as a function of the inhomogeneous layer's width, for values of the longitudinal wavenumber relevant for photospheric conditions. The results were found to match well the results found in another paper which studied the resistive slow surface sausage eigenmode. We also discuss the perturbation profiles of the quasimode and the eigenfunctions of continuum modes.}
   {}


   \maketitle
%

\section{Introduction}

Magnetohydrodynamic (MHD) waves are an ubiquitous phenomenon in the solar atmosphere. The observation of waves in the corona \citep{SchrijverEtAl1999, AschwandenEtAl1999, NakariakovEtAl1999, TomczykEtAl2007}, chromosphere \citep{DePontieuEtAl2007, MortonEtAl2012, Verth&Jess2016} and photosphere \citep{DorotovicEtAl2008, Fujimura&Tsuneta2009, GrantEtAl2015, MoreelsEtAl2015, KeysEtAl2018, Gilchrist-MillarEtAl2021} allows for both solar atmospheric seismology and the study of the coronal heating problem. 

The inference of plasma parameters by matching observed oscillations with theoretical results has been done both for the solar corona \citep{NakariakovEtAl1999, Nakariakov&Ofman2001, Nakariakov&Verwichte2005, AschwandenEtAl2003, GoossensEtAl2008, AndriesEtAl2005, VanDoorsselaereEtAl2011b} and the photosphere \citep{Fujimura&Tsuneta2009, Moreels&VanDoorsselaere2013}. Coronal structures such as loops and filament threads in particular are also known to harbor MHD waves which contribute to the heating of the corona \citep{Parnell&DeMoortel2012, Arregui2015, NakariakovEtAl2016, Nakariakov&Kolotkov2020, VanDoorsselaereEtAl2020}. More recently, observations in a photospheric pore of propagating slow surface sausage modes reported by \citet{GrantEtAl2015} were damped over a short enough length scale to be able to heat the chromosphere, suggesting waves in structures of the lower atmosphere are relevant to this problem as well. The energy the oscillations carry can be partially conveyed to the background plasma through various processes such as phase mixing, resonant absorption \citep{Zaitsev&Stepanov1975, Hollweg&Yang1988, HollwegEtAl1990, GoossensEtAl1992, CadezEtAl1997, ErdelyiEtAl2001, GoossensEtAl2002a, SolerEtAl2009c, SolerEtAl2013, YuEtAl2017b}, mode coupling \citep{PascoeEtAl2010, PascoeEtAl2012, HollwegEtAl2013, DeMoortelEtAl2016} and the development of turbulence by the Kelvin-Helmholtz instability \citep{Heyvaerts&Priest1983, OfmanEtAl1994, KarpenEtAl1994, KarampelasEtAl2017, AfanasyevEtAl2019, HillierEtAl2020, ShiEtAl2021, Geeraerts&VanDoorsselaere2021}.

When studying them theoretically through an analytical model, solar atmospheric structures such as coronal loops, filament threads, sunspots and photospheric pores are often modelled as a straight cylinder with a circular base in a first approximation. \citet{Roberts&Webb1979}, \citet{Wentzel1979}, \citet{Spruit1982}, and \citet{Edwin&Roberts1983}, among others, discussed cylinder modes in ideal MHD analytically for the case where the structure has a discontinuous boundary. This assumption of a discontinuous separation between two homogeneous plasmas inside and outside the cylinder is, however, a very crude approximation to reality. Indeed, the introduction of an inhomogeneous boundary layer gives rise to new physics, among which the process of resonant absorption. With the emergence of two continua in the spectrum of ideal MHD, namely the Alfv\'en and cusp continua, new local oscillations called continuum modes come into existence. These localized modes can be excited by either externally driven waves or a discrete eigenmode of the cylinder that couples to them, in a process called resonant absorption. In the case of an eigenmode being resonantly absorbed, it is damped because of its energy being transferred to local Alfv\'en or slow waves within the boundary layer. This happens at the position where the frequency of the eigenmode equals one of the continuum frequencies.

In the eigenvalue problem of linearized ideal MHD, where one assumes the waves to be normal modes of the system, the frequency of a discrete mode which is resonantly absorbed is complex because of the damping. However, the ideal MHD force operator being self-adjoint \citep{Goedbloed&Poedts2004}, its eigenvalues must be real and hence the resonantly absorbed discrete mode is not an eigenmode anymore. Instead, it becomes a so-called quasimode (also called a virtual eigenmode or a collective mode), which is the response of a system being excited at one of its natural frequencies and being damped because of its energy being transferred to a continuum mode. It is a dominant, global, exponentially decaying response to an initial perturbation \citep{Tirry&Goossens1996}, which cannot be distinguished from a true eigenmode in a time much smaller than the damping time of the mode and which has been discussed in detail for example by \citet{Sedlacek1971}, \citet{Zhu&Kivelson1988} and \citet{Goedbloed&Poedts2004} in different contexts. The efficiency of a quasimode in transferring heat to the surrounding plasma has been ascertained in works such as \citet{PoedtsEtAl1989}, \citet{Poedts&Kerner1991}, \citet{Tirry&Goossens1996}, \citet{DeGroof&Goossens2000} and \citet{Goossens&DeGroof2001}.

When resistivity is added in the model, the singularities in the differential equations giving rise to the continua disappear and the continuum modes become discrete eigenmodes instead \citep{GoedbloedEtAl2010}. The quasimode also becomes a true eigenmode now, as was discussed by \citet{Poedts&Kerner1991}. Its damping is then stronger, due to both resonant absorption and resistivity. The quasimode frequency is recovered in the limit of vanishing resistivity. As the current paper focuses on quasimodes with the oscillatory part of their frequency within the cusp continuum, photospheric conditions are of particular interest since then, in a cylinder with a discontinuous boundary, the slow surface mode has its frequency within the interval that becomes the cusp continuum when an inhomogeneous layer is included \citep{Edwin&Roberts1983}. The efficiency of damping due to resonant absorption of the slow surface mode in the cusp continuum compared to purely resistive damping has been studied by \citet{ChenEtAl2018} for the sausage mode and \citet{ChenEtAl2021} for the kink mode, in photospheric conditions. Their results suggest electrical resistivity is more efficient overall by about an order of magnitude, although damping due to resonant absorption of the slow surface kink mode in the cusp continuum dominates both damping due to resonant absorption in the Alfv\'en continuum and resistive damping at the lower end of the relevant values of resistivity.

In this paper, we work out an analytical method to find the dispersion relation of modes in a cylinder with an inhomogeneous layer of arbitrary width separating two homogeneous regions inside and outside. The method follows the work of \citet{SolerEtAl2013}, who used Frobenius series around the Alfv\'en resonant position to represent the solution of the eigenfunctions in the inhomogeneous layer. They assume the plasma to be pressureless, making the cusp continuum disappear from the system, and focus on kink modes resonantly absorbed in the Alfv\'en continuum in coronal conditions. The aim of the present paper is to extend their method to a plasma where thermal pressure is included, and to focus rather on slow surface modes resonantly absorbed in the cusp continuum. The results will not only allow us to understand the damping of the resulting quasimode and compare it to the numerical resistive results of \citet{ChenEtAl2018}, but also to study the behavior of the quasimode perturbations around the cusp resonant point. Additionally, the method can be used to plot the eigenfunctions of continuum modes.

\section{Model}\label{Model}

\begin{figure}
   \centering
   \includegraphics[scale=0.4]{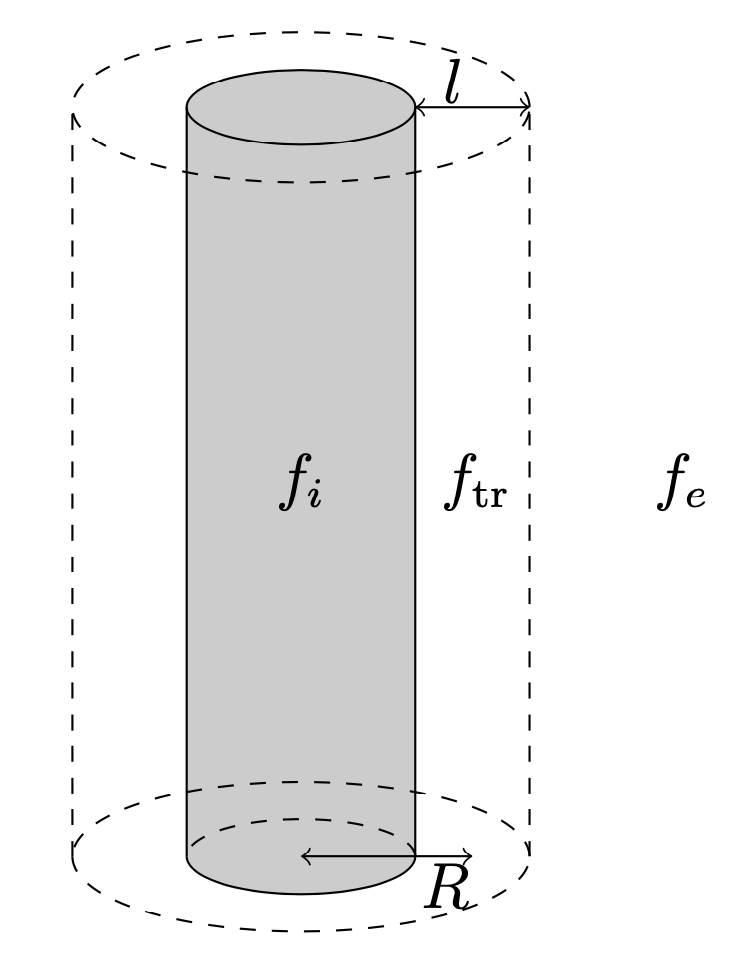}
      \caption{Sketch of the model cylinder. Quantities assume a constant value $f_i$ inside, a possibly different constant value $f_e$ outside, and have a continuously varying profile $f_{tr}$ linking the two through the inhomogeneous layer. The radius of the cylinder is denoted by $R$ and the layer width is denoted by $l$.}
         \label{sketch}
   \end{figure}

We consider a cylindrical solar atmospheric flux tube, having a circular cross section and an inhomogeneous transition layer at the boundary that separates two regions of homogeneous plasma with different properties (see Fig. \ref{sketch}). Although modes in photospheric structures such as pores and sunspots will be the main focus of this paper, the method we outline here can be repeated under different conditions.


In this model we will be working in the framework of linearized ideal MHD, and cylindrical coordinates $(r, \phi, z)$ will be used. The main physical quantities of interest are mass density, plasma velocity, magnetic field and thermal pressure, respectively denoted by $\rho$, $\pmb{v}$, $\pmb{B}$, and $p$. For a quantity $f$, its linearized form will be written as $f_0 + f_1$, where $f_0$ stands for the background part and $f_1$ for the first-order perturbation. We assume that there is no background velocity ($\pmb{v}_0 = \pmb{0}$), and that the background magnetic field is aligned with the cylinder's axis, as the pores and sunspots we have in mind typically have strong axial magnetic fields which act as a waveguide. We thus take $\pmb{B}_0 = B_{0z} \pmb{1}_z$. Since the background quantities are independent of $\p$, $z$ and $t$, the perturbed quantities can be Fourier-analyzed as $f_1 = \tilde{f}_1(r) \e$. In what follows we will drop the tilde and by $f_1$ we will actually mean $\tilde{f}_1$, for any quantity $f$.

The background quantities are assumed constant in both the internal and external regions, although the values are assumed to be possibly different in both regions. In the inhomogeneous layer at the boundary of the cylinder, that is to say for $r \in [R-l/2, R+l/2]$ with $R$ the cylinder radius and $l$ the layer width, the quantities are assumed to vary continuously in $r$ and to follow a predefined profile. It should be noted that, in reality, the profiles for the variation of the quantities are unknown and one thus has to make an arbitrary assumption about them in analytical models.

From the previously cited quantities, one can define the Lagrangian plasma displacement $\pmb{\xi}$ from $\pmb{v} = D_m(\pmb{\xi})$ (with $D_m$ denoting the material derivative), which equals $\frac{\partial \pmb{\xi}}{\partial t}$ in the linear case without background velocity, and the perturbed total pressure $P_1$ as $P_1 = p_1 + \frac{\pmb{B}_0 \cdot \pmb{B}_1}{\mu_0}$ (i.e., the sum of the perturbed thermal and magnetic pressures), with $\mu_0$ the magnetic permeability of free space. Following \citet{AppertEtAl1974}, the ideal MHD quantities can be reduced to two coupled first-order ordinary differential equations (ODE) for $\xi_r$ and $P_1$ under the assumptions previously mentioned :

\begin{align}
&D \dod{}{r} \l( r \xi_r \r) = -C_1 r P_1 \text{,} \label{D1}\\
&D \dod{P_1}{r} = C_2 \xi_r \text{,} \label{D2}
\end{align}
where 

\begin{align}
&D = \rho \l( v_A^2 + v_s^2 \r) \l( \o^2 - \o_A^2 \r) \l( \o^2 - \o_C^2 \r) \text{,}\\
&C_1 = \o^4 - \l(v_A^2 + v_s^2 \r) \l(\d\frac{n^2}{r^2} + k_z^2 \r) \l( \o^2 - \o_C^2 \r) \text{,}\\
&C_2 = \rho_0^2 \l( v_A^2 + v_s^2 \r) \l( \o^2 - \o_A^2 \r)^2 \l( \o^2 - \o_C^2 \r) \text{,}
\end{align}
and with $v_A = B_{0z}/\sqrt{\mu_0 \rho_0}$ the Alfvén speed, $v_s = \sqrt{\gamma p_0/\rho_0}$ the sound speed, $v_C = v_A v_s / (v_A^2 + v_s^2)^{1/2}$ the cusp speed, $\o_A = k_z v_A$ the Alfv\'en frequency, and $\o_C = k_z v_C$ the cusp frequency. Equations \eqref{D1} and \eqref{D2} can be combined into a single second-order ODE for $P_1$:

\begin{equation} \label{eqP1}
\dod[2]{P_1}{r} + \l\{ \frac{1}{r} - \frac{\dod{}{r} \l[ \rho_0 \l( \o^2 - \o_A^2 \r) \r]}{\rho_0 \l( \o^2 - \o_A^2 \r)} \r\} \dod{P_1}{r} + \l( -m^2 - \frac{n^2}{r^2} \r) P_1 = 0 \text{,}
\end{equation}
where 

\begin{equation}
m = \pm \sqrt{\d\frac{\l( \o_A^2 - \o^2 \r) \l( \o_s^2 - \o^2 \r)}{\l( \o_C^2 - \o^2 \r) \l( v_A^2 + v_s^2 \r)}} \text{,}
\end{equation}
and with $\o_s = k_z v_s$ the sound frequency. Here, $\sqrt{z}$ of a number $z \in \mathbb{C}$ is meant to be the solution $w \in \mathbb{C}$ to $z=w^2$ with $-\frac{\pi}{2} < \text{Arg}(w) \leq \frac{\pi}{2}$. Then, from Eq. \eqref{D2}, the expression for $\xi_r$ can be derived once $P_1$ is known:

\begin{equation} \label{eqXir}
\xi_r = \d\frac{1}{\rho_0 \l( \o^2 - \o_A^2 \r)} \dod{P_1}{r} \text{.}
\end{equation}

\section{Finding the solutions for $P_1$ and $\xi_r$} \label{Solutions}

In this section, we seek a solution for $P_1$ from Eq. \eqref{eqP1}. Knowing the solution of $P_1$, one can determine the solution of $\xi_r$ from Eq. \eqref{eqXir}. The solutions for the other quantities can then be derived from $P_1$ and $\xi_r$. There are three regions where different solution forms will occur: inside the cylinder, in the inhomogeneous layer, and outside the cylinder.

\subsection{Solutions in the internal and external regions}

In the homogeneous plasmas of the internal (i.e., where $r< R-l/2$) and external (i.e., where $r>R+l/2$) regions, Eq. \eqref{eqP1} will simplify because all background quantities are constant. Indeed, the ODE for $P_1$ will reduce to 

\begin{equation} \label{eqP1ConstBack}
\dod[2]{P_1}{r} + \d\frac{1}{r} \dod{P_1}{r} + \l( -m^2 - \frac{n^2}{r^2} \r) P_1 = 0 \text{.}
\end{equation}
This is a Bessel equation, the solution of which is well known and can be found for example in \citet{Edwin&Roberts1983}. For surface modes, the internal solution $P_{1i}$ and the external solution $P_{1e}$ are then

\begin{align}
&P_{1i} = C_1 I_n(m_i r) \text{,} \label{SolP1Homogi}\\
&P_{1e} = C_2 K_n( m_e r) \text{,} \label{SolP1Homoge}
\end{align}
where $I_n$ is the modified Bessel function of the first kind of order $n$, $K_n$ is the modified Bessel function of the second kind of order $n$, $m_i$ and $m_e$ are the versions of $m$ in respectively the internal and external regions, whereas $C_1$ and $C_2$ are constants. In what follows we will need the solutions for $\xi_r$ as well, which are obtained from Eq. \eqref{eqXir} and are given as follows for the internal and external regions:

\begin{align}
&\xi_{ri} = \d\frac{C_1 m_i}{\rho_{0i} \l( \o^2 - \o_{Ai}^2 \r)} I_n'(m_i r) \text{,}\\
&\xi_{re} = \d\frac{C_2 m_e}{\rho_{0e} \l( \o^2 - \o_{Ae}^2 \r)} K_n'(m_e r) \text{.}
\end{align}

\subsection{Solutions in the inhomogeneous layer} \label{SolInh}

In the inhomogeneous boundary layer, the background quantities are not constant and vary continuously from their value inside the cylinder to their value in the surrounding plasma. Therefore, the full ODE for $P_1$, Eq. \eqref{eqP1}, needs to be solved. The radial variation of the background quantities in the boundary layer, specifically $v_A$ and $v_C$, gives rise to singularities of the ODE \eqref{eqP1} at positions $r_A$ and $r_C$ which respectively satisfy $\o^2 = \o_A^2(r_A)$ and $\o^2 = \o_C^2(r_C)$. These singularities are both regular singular points of the ODE and were discussed by \citet{SakuraiEtAl1991} in the limit of a thin layer (i.e., $l/R \ll 1$) where they were assumed to be real. In what follows, we will assume that $v_A^2$ and $v_C^2$ are strictly monotonic functions of $r$.

To solve Eq. \eqref{eqP1}, we will work out a method which is based on the one developed by \citet{SolerEtAl2013}. The authors of that paper use Frobenius series solutions around the Alfv\'en resonant point, in order to represent the solution of $P_1$ inside the layer. With the inclusion of a cusp resonance point, however, complications arise compared to the case discussed by \citet{SolerEtAl2013}. Indeed, because of the presence of both Alfv\'en and cusp resonance, there will now be two singularities arising within the inhomogeneous layer. A Frobenius series around one of the resonant points will have its convergence radius limited because of the presence of the other resonant point nearby, and vice versa. This means the whole layer will not be covered by the convergence disc of a single series when both resonant points are present. In addition, depending on the transition profiles chosen for the background quantities in the inhomogeneous layer, additional singularities of the coefficients in the ODE \eqref{eqP1} may be present near the layer and also have an impact on the convergence radii of the series. We will thus need to make multiple series expansions: around the cusp resonant position ($r_C$), the Alfv\'en resonant position ($r_A$), and some regular points of Eq. \eqref{eqP1} until a whole path in the complex $r$-plane which links the points $R-l/2$ and $R+l/2$ is covered by convergence discs.

\subsubsection{Frobenius solution around $r_C$} \label{FrobrC}

The resonant position $r_C$ being a regular singular point of the ODE \eqref{eqP1} as well \citep{SakuraiEtAl1991}, we write a solution locally around $r_C$ as a Frobenius series:

\begin{equation}
P_1(r) = \d\sum_{k=0}^{\infty} p_k \l( r-r_C \r)^{k+s} \text{.}
\end{equation}
By inserting this solution into the equation, one finds the indicial equation. \citet{SakuraiEtAl1991} showed that this equation is $s (s-1)= 0$ for a strictly monotonic transition profile of $v_A^2$. Hence two basic independent solutions of Eq. \eqref{eqP1} are

\begin{align}
&P_{1,1,C}(r) = \d\sum_{k=0}^{\infty} \alpha_k \l(r-r_C \r)^{k+1} \text{,} \label{P11C}\\
&P_{1,2,C}(r) = \d\sum_{k=0}^{\infty} \sigma_k \l(r-r_C \r)^{k} + \mathcal{C}_C P_{1,1,C}\l(r-r_C \r) \ln\l(r-r_C \r) \text{,} \label{P12C}
\end{align}
where the coefficients $\alpha_0$ and $\sigma_0$ can both be freely chosen because of the two degrees of freedom in the general solution of a second-order ODE, whereas the remaining $\alpha_k$ and $\sigma_k$, as well as the constant $\mathcal{C}_C$, are determined by recursion relations (see Appendix \ref{Recursions}). From now on, $\alpha_0$ and $\sigma_0$ will both be taken equal to $1$. The logarithm in the solution \eqref{P12C} appears because of the cusp resonance, as will be clarified later. The general solution is then given by

\begin{equation} \label{P1C}
P_1(r) = A_1 P_{1,1,C}(r) + A_2 P_{1,2,C}(r)
\end{equation}
within the convergence radius of the two involved power series, with $A_1$ and $A_2$ some arbitrary coefficients. 

The expression for the quantity $\mathcal{C}_C$ in Eq. \eqref{P12C} is determined by the recursion relation for the coefficients $\sigma_k$ and will be useful later on. In Appendix \ref{AppC_C}, we show that it is given by

\begin{equation} \label{C_C}
\mathcal{C}_C = -\d\frac{k_z^4 v_C^4(r_C)}{\l(v_A^2(r_C) + v_s^2(r_C) \r) \tod{\o_C^2}{\z}\Bigr\rvert_{r=r_C}} \text{.}
\end{equation}

\subsubsection{Frobenius solution around $r_A$}

For the solution of Eq. \eqref{eqP1} in the form of a Frobenius series around the Alfv\'en resonant position $r_A$, we proceed similarly by inserting

\begin{equation}
P_1(r) = \d\sum_{k=0}^{\infty} p_k \l( r-r_A \r)^{k+s} \text{.}
\end{equation}
into the ODE. The obtained indicial equation in this case is $s(s-2)=0$, as was also derived by \citet{SakuraiEtAl1991} for a strictly monotonic transition profile of $v_A^2$. Hence, two independent basic solutions are

\begin{align}
&P_{1,1,A}(r) = \d\sum_{k=0}^{\infty} \beta_k \l(r-r_A \r)^{k+2} \text{,}\\
&P_{1,2,A}(r) = \d\sum_{k=0}^{\infty} \tau_k \l(r-r_A \r)^{k} + \mathcal{C}_A P_{1,1,A}\l(r-r_A \r) \ln\l(r-r_A \r) \text{,} \label{P12}
\end{align}
where the coefficients $\beta_0$ and $\tau_0$ are chosen freely, whereas the remaining $\beta_k$ and $\tau_k$ as well as the constant $\mathcal{C}_A$ are determined by recursion relations (see Appendix \ref{Recursions}). From now on, $\b_0$ and $\t_0$ will both be taken equal to $1$ as well. The logarithm in the solution \eqref{P12} appears because of the Alfv\'en resonance, as will also be clarified later. The general solution is then given by

\begin{equation} \label{P1A}
P_1(r) = A_3 P_{1,1,A}(r) + A_4 P_{1,2,A}(r)
\end{equation}
within the convergence radius of the two involved power series, with $A_3$ and $A_4$ some arbitrary coefficients. The value of the quantity $\mathcal{C}_A$ is determined by the recursive relation for the $\tau_k$.

\subsubsection{Power series solution around regular points}

The power series solution around a regular point $r_0$ of the ODE \eqref{eqP1} can be found by inserting a general power series of the form

\begin{equation} \label{P1r0theo}
P_1(r) = \d\sum_{k=0}^{\infty} p_k \l( r-r_0 \r)^k
\end{equation}
in Eq. \eqref{eqP1}. The two first coefficients can be chosen freely, yielding the two basic solutions

\begin{align}
&P_{1,1,r_0}(r) = \d\sum_{k=0}^{\infty} a_k \l( r-r_0 \r)^k \text{,} \label{P11r0}\\
&P_{1,2,r_0}(r) = \d\sum_{k=0}^{\infty} b_k \l( r-r_0 \r)^k \text{,} \label{P12r0}
\end{align}
which are independent if the couples $(a_0, a_1)$ and $(b_0, b_1)$ are not multiples of one another. The remaining $a_k$ and $b_k$ are again determined by a recursion relation (see Appendix \ref{Recursions}). The general solution is then given by

\begin{equation} \label{P1r0}
P_1(r) = A_{r_0} P_{1,1,r_0}(r) + B_{r_0} P_{1,2,r_0}(r)
\end{equation}
within the convergence radius of the two involved power series, with $A_{r_0}$ and $B_{r_0}$ some arbitrary coefficients.

\section{Constructing the dispersion relation}

To find the eigenmodes and quasimodes of the system, one needs to find a dispersion relation which expresses the frequency $\o$ as a function of the wavenumbers $k_z$ and $n$. To achieve this, boundary conditions need to be imposed at the positions $r_i = R-l/2$ and $r_e=R+l/2$ where the internal and external solutions are linked with the solution in the inhomogeneous layer.

\begin{figure}
   \centering
   \includegraphics[scale=0.18]{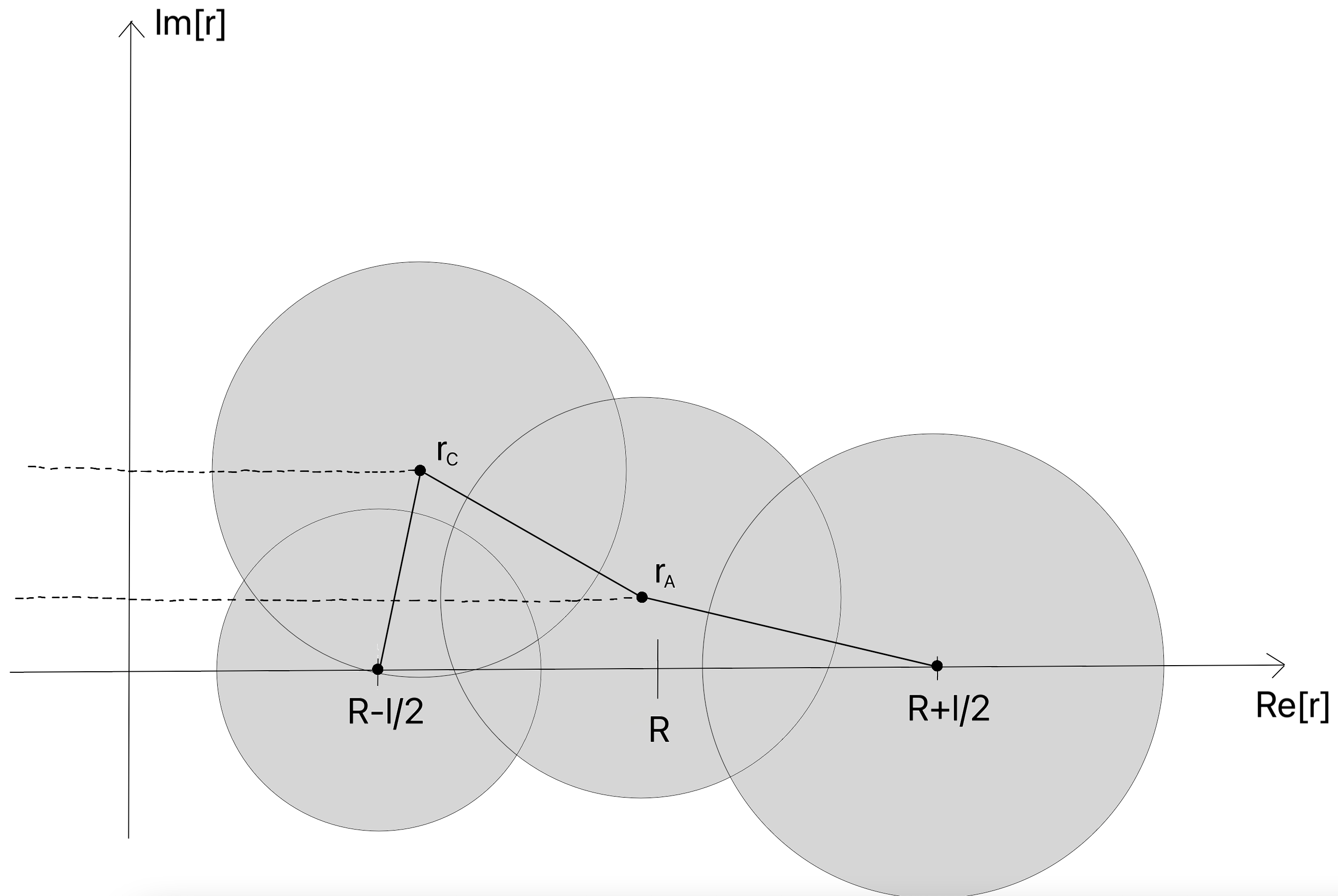}
      \caption{A representation of a path (thick solid line) linking $r_i=R-l/2$ with $r_e=R+l/2$, formed by the line segments $[r_i r_C]$, $[r_C r_A]$, and $[r_A r_e]$. The dashed lines represent branch cuts due to the presence of the logarithms in the Frobenius solutions. In this representation, the represented discs only have radii of $80$\% of those of the corresponding convergence discs, as this was used in numerical calculations to avoid inaccuracy due to the slow convergence of the series near the edges of the convergence discs.}
         \label{path}
   \end{figure}

In order to satisfy the boundary conditions at the points $r_i$ and $r_e$, we need to connect them with a path which is fully covered by convergence discs of series expansions around points in the complex $r$-plane (see Fig. \ref{path}), following a process called analytic continuation. The effects of both the cusp and the Alfv\'en resonances obviously need to be taken into account, so both $r_C$ and $r_A$ must lie on the path. We choose the path such that the coefficients of Eq. \eqref{eqP1} are analytic in all other points composing it, so that we can locally represent the solution of the ODE as regular power series around those points. In order to do this, we first identify the singularities of the coefficients of Eq. \eqref{eqP1}. These depend on the transition profiles chosen for the background quantities in the inhomogeneous layer. It is assumed that $r_i$ and $r_e$ are analytic points of the coefficients of Eq. \eqref{eqP1}, as will be the case in the example discussed in Section \ref{Quasim}.

As can be seen from the Frobenius solutions in the previous section, both $r_C$ and $r_A$ are logarithmic branch points. The path may in principle not cross branch cuts, as the series expansions would then be defining $P_1$ on the next Riemann sheet. Theoretically, it would be sufficient to make the path circumvent a branch point and its branch cut, but in practice we found that in some situations substantial numerical accuracy is lost in this way due to a high number of expansion points, as well as convergence discs with a very small radius being needed. Instead, we always choose the shortest path, and, when needing to cross a branch cut, we correct for the jump at the last encountered branch point.

\subsection{Nonoverlapping cusp and Alfv\'en continua} \label{nonoverlapping}

\citet{SakuraiEtAl1991} derived connection formulae for both $P_1$ and $\xi_r$ across the cusp resonant layer in the thin boundary (TB) limit (i.e., $l/R \ll 1$). In the case where the cusp and Alfv\'en continua are not overlapping, these can be used to derive the dispersion relation for modes that undergo resonant absorption in the cusp continuum. This situation has been discussed for example by \citet{YuEtAl2017b}, who derived the corresponding dispersion relation in that limit (see Eq. (27) therein). That dispersion relation can be recovered from our model, as we will show.
 
\subsubsection{General dispersion relation} \label{GenDispNonov}

If we assume that the cusp and Alfv\'en continua do not overlap, then only one resonant point will be present at a time. To find a dispersion relation in this particular case, we compute the series expansions of two basic independent solutions $P_{1,1}$ and $P_{1,2}$ of the ODE \eqref{eqP1} along a path in the complex $r$-plane which links $r_i$ with $r_e$ and includes $r_C$. Such solutions are defined over a subset of $\mathbb{C}$, but are locally represented by a power series (if expanding around a regular point) or a Frobenius solution (if expanding around a regular singular point). The two basic solutions $P_{1,1}$ and $P_{1,2}$ we choose are the ones that are locally represented respectively by the Frobenius solutions $P_{1,1,C}$ and $P_{1,2,C}$ from Eqs. \eqref{P11C} and \eqref{P12C}, with $\a_0 = \s_0 = 1$ within the convergence disc of the involved power series. For each of the two solutions $P_{1,1}$ and $P_{1,2}$, we then find the local representation as a power series or Frobenius solution around a new point on the path and within the convergence disc of the former local representation, and so on until we covered the whole path. 

For this, we use the following result: two power series $\sum_{k=0}^{\infty} f_k (r-r_1)^k$ and $\sum_{k=0}^{\infty} g_k (r-r_2)^k$ are equal on the overlapping region between their respective convergence discs if

\begin{equation} \label{movingseries}
g_k = \d\sum_{j=k}^{\infty} f_j \binom{j}{k} \l( r_2-r_1 \r)^{j-k} \text{.}
\end{equation}
We thus calculate the first two coefficients of the new power series expansion with formula \eqref{movingseries}, and compute the remaining ones from the recursion relation for the series expansion \eqref{P1r0theo} around a regular point.

Once we have series representations for the complex functions $P_{1,1}$ and $P_{1,2}$ along the whole path, we can compute $P_{1,1}(r_i)$, $P_{1,1}(r_e)$, $P_{1,2}(r_i)$, and $P_{1,2}(r_e)$. Denoting by $[f]_i$ and $[f]_e$ the jumps of a quantity $f$ across respectively the ends $r_i$ and $r_e$ of the inhomogeneous layer, we impose the following four boundary conditions which need to be satisfied for physical reasons:

\begin{equation} \label{BCs}
[P_1]_i = 0 \text{, \quad} [P_1]_e = 0 \text{, \quad} [\xi_r]_i = 0 \text{, \quad} [\xi_r]_e = 0 \text{.}
\end{equation}
Introducing the notations

\begin{align}
&\mathcal{G}_{i,e} = P_{1,1}(r_{i,e})\\
&\mathcal{F}_{i,e} = P_{1,2}(r_{i,e})\\
&\Xi_{i,e} = \d\frac{1}{\rho_{0}(r_{i,e}) \l(\o^2 - \o_{A}^2(r_{i,e}) \r)} \dod{P_{1,1}}{r}(r_{i,e})\\
&\Gamma_{i,e} = \d\frac{1}{\rho_{0}(r_{i,e}) \l(\o^2 - \o_{A}^2(r_{i,e}) \r)} \dod{P_{1,2}}{r}(r_{i,e}) \text{,}
\end{align}
we can then rewrite Eqs. \eqref{BCs} as

\begin{align}
C_1 I_n(m_i r_i) &= C_3 \mathcal{G}_i + C_4 \mathcal{F}_i \text{,} \label{BC1} \\
C_2 K_n(m_e r_e) &= C_3 \mathcal{G}_e + C_4 \mathcal{F}_e \text{,} \\
\d\frac{C_1 m_i}{\rho_{0i} \l(\o^2 - \o_{Ai}^2 \r)} I_n'(m_i r_i) &= C_3 \Xi_i + C_4 \Gamma_i \text{,} \\
\d\frac{C_2 m_e}{\rho_{0e} \l(\o^2 - \o_{Ae}^2 \r)} K_n'(m_e r_e) &= C_3 \Xi_e + C_4 \Gamma_e \text{.} \label{BC4}
\end{align}
Equations \eqref{BC1}-\eqref{BC4} form a system of four equations in the four unknowns $C_1$, $C_2$, $C_3$, and $C_4$. In order to have a nontrivial solution, the determinant of this system needs to vanish. This yields the following dispersion relation:

\begin{equation} \label{Disp}
\d\frac{\frac{m_e}{\rho_{0e} \l( \o^2 - \o_{Ae}^2 \r)} \frac{K_n'(m_e r_e)}{K_n(m_e r_e)} \mathcal{G}_e - \Xi_e}{\frac{m_e}{\rho_{0e} \l( \o^2 - \o_{Ae}^2 \r)} \frac{K_n'(m_e r_e)}{K_n(m_e r_e)} \mathcal{F}_e - \Gamma_e} - \frac{\frac{m_i}{\rho_{0i} \l( \o^2 - \o_{Ai}^2 \r)} \frac{I_n'(m_i r_i)}{I_n(m_i r_i)} \mathcal{G}_i - \Xi_i}{\frac{m_i}{\rho_{0i} \l( \o^2 - \o_{Ai}^2 \r)} \frac{I_n'(m_i r_i)}{I_n(m_i r_i)} \mathcal{F}_i - \Gamma_i} = 0 \text{.}
\end{equation}

The left-hand side of this equation will be called the dispersion function. It is multivalued because of the logarithm $\ln(r-r_C)$ in $P_{1,2}$, so one needs to choose where to lay the branch cut. We will take it to lie on the negative real axis of $r-r_C$, in which case the branch cuts of the dispersion function will lie exactly on the cusp continuum when viewed as a function of $\o$. The reason for making this choice will be explained later.

\subsubsection{Thin boundary limit}

In the TB limit, that is to say when $l/R \ll 1$, the dispersion relation \eqref{Disp} can be approximated. In order to do so, we note that, in this limit, $r \approx r_C$ for every $r$ within the inhomogeneous layer since it is narrow. We will therefore make the approximation that $r_C \approx R$. The series expansions in Eqs. \eqref{P11C}, \eqref{P12C}, \eqref{P11r0}, and \eqref{P12r0} can then be approximated by only including the zeroth-order terms of the respective series expansions, yielding the following approximations:

\begin{align}
&\mathcal{G}_{i,e} \approx 0 \text{,}\\
&\mathcal{F}_{i,e} \approx 1 \text{,}\\
&\Xi_{i,e} \approx \d\frac{1}{\rho_{0}(R) \l(\o_C^2(R) - \o_{A}^2(R) \r)} \text{,}\\
&\Gamma_{i,e} \approx \frac{\mathcal{C}_C \l( 1 + \ln (\mp l/2) \r)}{\rho_{0}(R) \l(\o_C^2(R) - \o_{A}^2(R) \r)} \text{.}
\end{align}
Then, making the approximations $m_e r_e \approx m_e R$ and $m_i r_i \approx m_i R$ in the arguments of the modified Bessel functions, and using the expression for $\mathcal{C}_C$ derived in Eq. \eqref{C_C}, the dispersion relation \eqref{Disp} is approximated by

\begin{align} \label{DispTB}
\d\frac{m_e}{\rho_{0e} \l( \o^2 - \o_{Ae}^2 \r)} \frac{K_n'(m_e R)}{K_n(m_e R)} - &\frac{m_i}{\rho_{0i} \l( \o^2 - \o_{Ai}^2 \r)} \frac{I_n'(m_i R)}{I_n(m_i R)} \notag\\
& \hspace{1.2cm} = -\frac{\pi v_C^4(R) \ln(-1)}{\rho_0(R) v_A^4(R) \tod{v_C^2}{r}\Bigr\rvert_{r=R}} \text{.}
\end{align}
Here, $\ln(-1)$ must be chosen as either $i \pi$ or $-i \pi$ such that the frequency of the mode has a negative imaginary part (corresponding to a damped mode). Equation \eqref{DispTB} is equivalent to the dispersion relation derived by \citet{YuEtAl2017b} in the thin boundary limit.

\subsection{Overlapping cusp and Alfv\'en continua}

In some situations, such as the photospheric conditions of \citet{Edwin&Roberts1983}, the value of the equilibrium quantities are such that the cusp and Alfv\'en continua overlap. For $n \neq 0$, this has the effect of introducing an Alfv\'en resonant position along with the cusp resonant position in the inhomogeneous layer. The sausage modes ($n = 0$) are not resonantly absorbed in the Alfv\'en continuum in this model \citep{SakuraiEtAl1991}, and hence for them there is only the resonant position $r_C$.

The method which we outlined in the previous section is useful for finding the general expression \eqref{Disp} for the dispersion relation, allowing us to recover the analytical approximation for non-overlapping continua in the thin boundary limit from \citet{YuEtAl2017b}. However, solving the dispersion relation in the general case needs to be done numerically, which is not done efficiently with this method. We also note that \citet{SolerEtAl2009c} found an analytical thin boundary approximation to the dispersion relation for overlapping continua, based on the individual jumps at the Alfv\'en and cusp resonant positions derived by \citet{SakuraiEtAl1991}. This relation will not be recovered here.

Firstly, we encounter an additional difficulty when the cusp and Alfv\'en continua are overlapping. Indeed, when trying to find local representations of the basic solutions $P_{1,1}$ and $P_{1,2}$ as either power series or Frobenius solutions, the Alfv\'en resonant position will need to be included on the path linking $r_i$ with $r_e$ along which we seek these representations. When making the calculations, we find that the presence of the nonzero constant $\mathcal{C}_A$ when $n \neq 0$ does not allow the method outlined before to be used. Secondly, we lose some accuracy with every new expansion point included in the path to analytically continue each of the two solutions $P_{1,1}$ and $P_{1,2}$. Some accuracy is lost for two reasons: by every series needing to be truncated, and by the first two coefficients of the series in every new local representation being approximations themselves due to the series in formula \eqref{movingseries} needing to be truncated as well. In addition, each new expansion point increases the computational cost of the numerical algorithm to create and solve the dispersion relation. This means that we need to minimize the number of local representations of a solution to the ODE \eqref{eqP1} in order to maximize efficiency in the numerical computations. 

A more practical method, which reduces both the number of local representations of solutions and the loss in accuracy in the computation of each representation, consists of covering the path with local representations of unrelated general solutions and linking them together with additional boundary conditions. The values of the first free coefficients in a series determine a specific solution. In the former method, we started from a specific solution and calculated the local representations of that solution along a path. We did this independently for two solutions, namely the one represented in a region around $r_C$ by the Frobenius solution \eqref{P11C} with coefficient $\a_0 = 1$ and the one represented in a region around $r_C$ by the Frobenius solution \eqref{P12C} with coefficients $\s_0=1$ and $\a_0 = 1$. For each new local representation, we thus needed to compute the corresponding values of the first free coefficients. The two solutions, being independent, together form a general solution \eqref{P1C} in the inhomogeneous layer. We then linked this general solution with the solutions outside the homogeneous layer through the boundary conditions \eqref{BCs}. In the following method, we instead compute local representations of general solutions of the form \eqref{P1C}, \eqref{P1A}, or \eqref{P1r0}, with coefficients $\a_0=1$ and $\s_0=1$ in the case of a Frobenius solution \eqref{P1C}, $\b_0=1$ and $\t_0=1$ in the case of a Frobenius solution \eqref{P1A}, and $(a_0, a_1) = (1,0)$ and $(b_0, b_1) = (0,1)$ in the case of a power series solution \eqref{P1r0}. This we do along a path in the complex $r$-plane which links $r_i$ with $r_e$, includes both $r_C$ and $r_A$, and is entirely covered by convergence discs around expansion points.

Each of these local representations then includes two arbitrary constants. On each region where the convergence discs of two neighboring general solutions overlap, we equate the general solutions of $P_1$ and $\xi_r$. This determines two out of the four involved constants. With $N$ expansion points covering the path, there will be $2N$ arbitrary constants involved. Equating neighboring general solutions of both $P_1$ and $\xi_r$ on every overlapping region between pairs of convergence discs yields $2N-2$ equations. This means $2N-2$ of the arbitrary constants can be written as a function of two remaining ones, which we will call $C_3$ and $C_4$. These two constants $C_3$ and $C_4$, which are related to the solution in the inhomogeneous layer, will then, together with the two constants $C_1$ and $C_2$ related to the solutions in the homogeneous regions respectively inside and outside the cylinder, yield a system of four equations in the four unknowns $C_1$, $C_2$, $C_3$ and $C_4$ when we impose the four boundary conditions \eqref{BCs}. For the obtained system to have nontrivial solutions, its $4 \times 4$ determinant must equal $0$. This determinant, when viewed as a function of the complex variable $\o$, is called the dispersion function. Equating the dispersion function to $0$ forms the dispersion relation, which needs to be solved numerically. It is this method that we use, rather than the one outlined in the previous section, in order to find the modes in the general case.

We note that the dispersion function is multivalued because of both $\ln(r-r_C)$ and $\ln(r-r_A)$. For general $n$ it will have two branch points $r_C$ and $r_A$ in the complex $r$-plane, and hence also two branch cuts (see Fig. \ref{path}). These branch cuts are arbitrary from a mathematical point of view, but from a physical standpoint they have to be taken such that the principal Riemann sheet of the dispersion function does not have complex zeros. The reason for this will be explained in the next section. It turns out that taking the branch cuts of both $\ln(r-r_C)$ and $\ln(r-r_A)$ to lie on the negative real axis of their respective arguments ensures that there are no complex zeros on the principal Riemann sheet of the dispersion function. Viewed in the complex $\o$ domain, the branch cuts of the dispersion function then correspond exactly to the cusp and Alfv\'en continua on the real axis. This also means that, in the case the continua overlap, the two branch cuts will also overlap. A path which crosses this overlapping region will then continue on one of two possible Riemann sheets.

\section{Frequency and perturbation profiles of the quasimode in photospheric conditions} \label{Quasim}

\subsection{Theoretical considerations}

In this section, we apply the methods outlined in the previous section to the specific case of a cylindrical structure in the photosphere, for example a pore or a sunspot. For this, we take $(v_{Ae}, v_{si}, v_{se}) = (1/4,1/2,3/4) v_{Ai}$, corresponding to the photospheric conditions in Fig. 3 of \citet{Edwin&Roberts1983}.

The existence of an inhomogeneous boundary layer results in the possibility for continuum modes with a frequency in the cusp continuum $[\o_{Ce}, \o_{Ci}]$ to be excited. Since  in the absence of a layer the slow surface mode has its frequency in that interval as well, it is natural to expect that this mode will couple to a local cusp continuum mode with the same frequency if a layer is present. The damping it undergoes because of the transfer of its energy to the continuum mode is then expected to be expressed in the imaginary part of its frequency becoming strictly negative, and the frequency thus becoming complex. 

However, it is a proven result that the ideal MHD operator is Hermitian and can thus only have real eigenfrequencies \citep{Goedbloed&Poedts2004}. Our dispersion relation being satisfied for some value $\o_0$ of the frequency is equivalent to finding a zero of the dispersion function at $\o_0$ on its principal Riemann sheet. Therefore, no complex zero of the dispersion function should be found on that sheet. By taking the branch cuts in the way we defined them in the previous section, this is ensured. 


The proper study of a quasimode, being a natural oscillatory response to an initial perturbation, needs to be done by solving the initial value problem with the Laplace transform. The quasimode is then found as a pole of a (multivalued) Green's function, namely as a zero on the neighbouring Riemann sheet of its denominator. This denominator is the dispersion function. The contribution of that pole is then taken into account when computing the inverse Laplace transform, by deforming the Bromwich countour across the branch cut (which is the continuum) onto the next Riemann sheet.

The method we developed in the present paper solves the eigenvalue problem by considering normal modes and constructing the dispersion function for eigenmodes from the physical considerations expressed in the boundary conditions \eqref{BCs}. The quasimode, not being an eigenmode, does hence not naturally appear when solving this problem. However, the dispersion function having the same zeros as the one in the initial value problem, one can find the frequencies of quasimodes by looking on its next Riemann sheet. The quasimode frequency $\o_0$ that we find there does not satisfy the dispersion relation, since on the principal Riemann sheet $\o_0$ is not a zero of the dispersion function. This translates in the fact that a quasimode perturbation cannot be continuous in the two boundaries of the inhomogeneous layer and at the resonant position at the same time. In fact, in order to access another Riemann sheet of the dispersion function, a different branch of the logarithm in the large Frobenius solution around a resonant position $r_{\text{res}}$ must be used on each side of the line $r=\text{Re}[r_{\text{res}}]$. This ensures the discontinuity occurs at the resonant position rather than at one of the boundaries of the layer, which is the only physically acceptable solution.

The aim of our method is to find the values of the complex frequency of the quasimode corresponding to the slow surface mode of the discontinuous boundary case in photospheric conditions, in order to quantify its damping. In particular, we would like to know the profile of the damping time as a function of the layer width $l$ for values of the longitudinal wavenumber $k_z$ which are realistic for oscillations in photospheric pores. It will also be interesting to plot the profiles of the quasimode perturbations, and compare them for different values of $l$ and $k_z$.

\subsection{Result for an example case} 

The dispersion function obviously depends on the transition profiles of the background quantities in the inhomogeneous layer. Theoretically, any profile could be taken. However, in the specific example we are now going to work out, we choose relatively simple transition profiles in order for the numerical calculations to remain feasible: we assume the squared cusp and squared sound speeds to have a linear profile in the layer. These in turn fix the profiles of the square Alfv\'en speed, magnetic field, thermal pressure and density. 

The linear profiles for the squared cusp and squared sound speeds in the inhomogeneous layer are taken as follows:

\begin{align}
&v_C^2(r) = \widetilde{v_C^2} r + \widehat{v_C^2} \text{,} \label{v_C^2Linear}\\
&v_s^2(r) = \widetilde{v_s^2} r + \widehat{v_s^2} \text{,}
\end{align}
with the constants defined as

\begin{align}
&\widetilde{v_C^2} = \d\frac{v_{Ce}^2 - v_{Ci}^2}{l} \text{,}\\
&\widetilde{v_s^2} = \d\frac{v_{se}^2 - v_{si}^2}{l} \text{,}
\end{align}
\begin{align}
&\widehat{v_C^2} = \l(\d\frac{1}{2} + \frac{R}{l} \r) v_{Ci}^2 + \l(\d\frac{1}{2} - \frac{R}{l} \r) v_{Ce}^2 \text{,}\\
&\widehat{v_s^2} = \l(\d\frac{1}{2} + \frac{R}{l} \r) v_{si}^2 + \l(\d\frac{1}{2} - \frac{R}{l} \r) v_{se}^2 \text{.}\label{v_s^2hat}
\end{align}
With these profiles, we can now try to find the quasimode which corresponds to the slow surface eigenmode in the discontinuous boundary case as a complex zero of the dispersion function on its first Riemann sheet neighboring the principal sheet. 

\subsubsection{Sausage mode}

\begin{figure}
   \centering
   \includegraphics[scale=0.28]{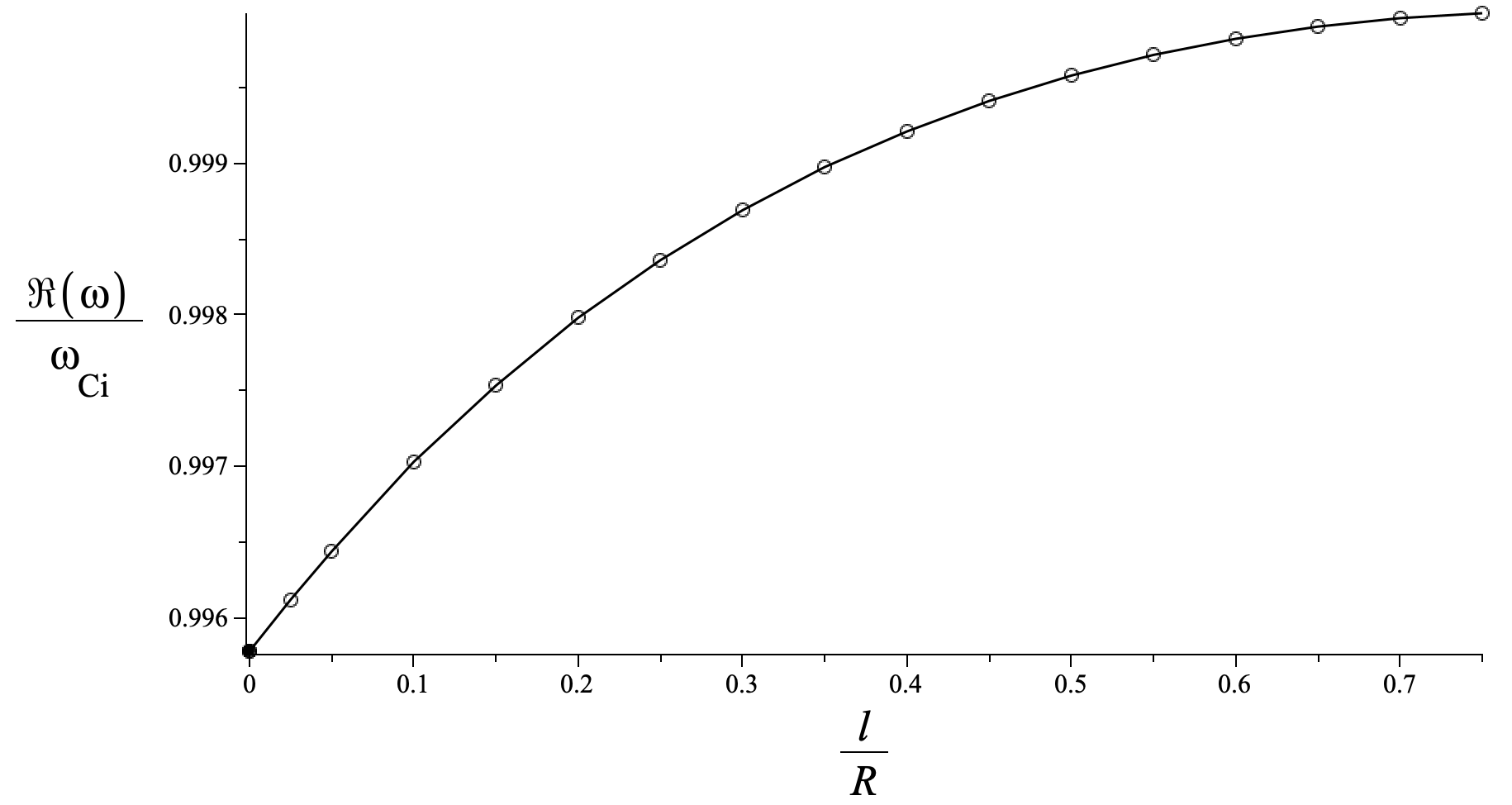}
   \includegraphics[scale=0.28]{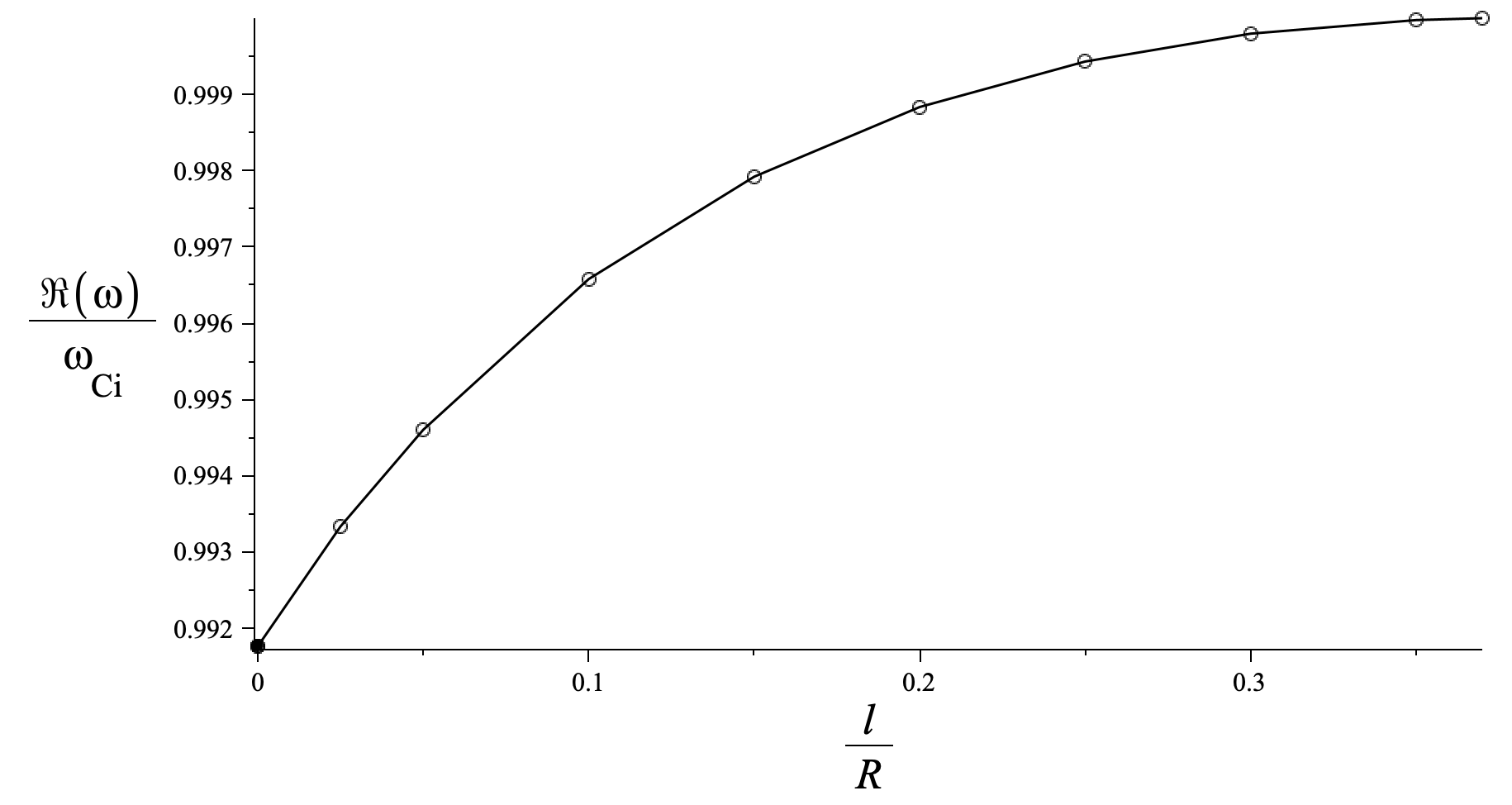}
   \includegraphics[scale=0.28]{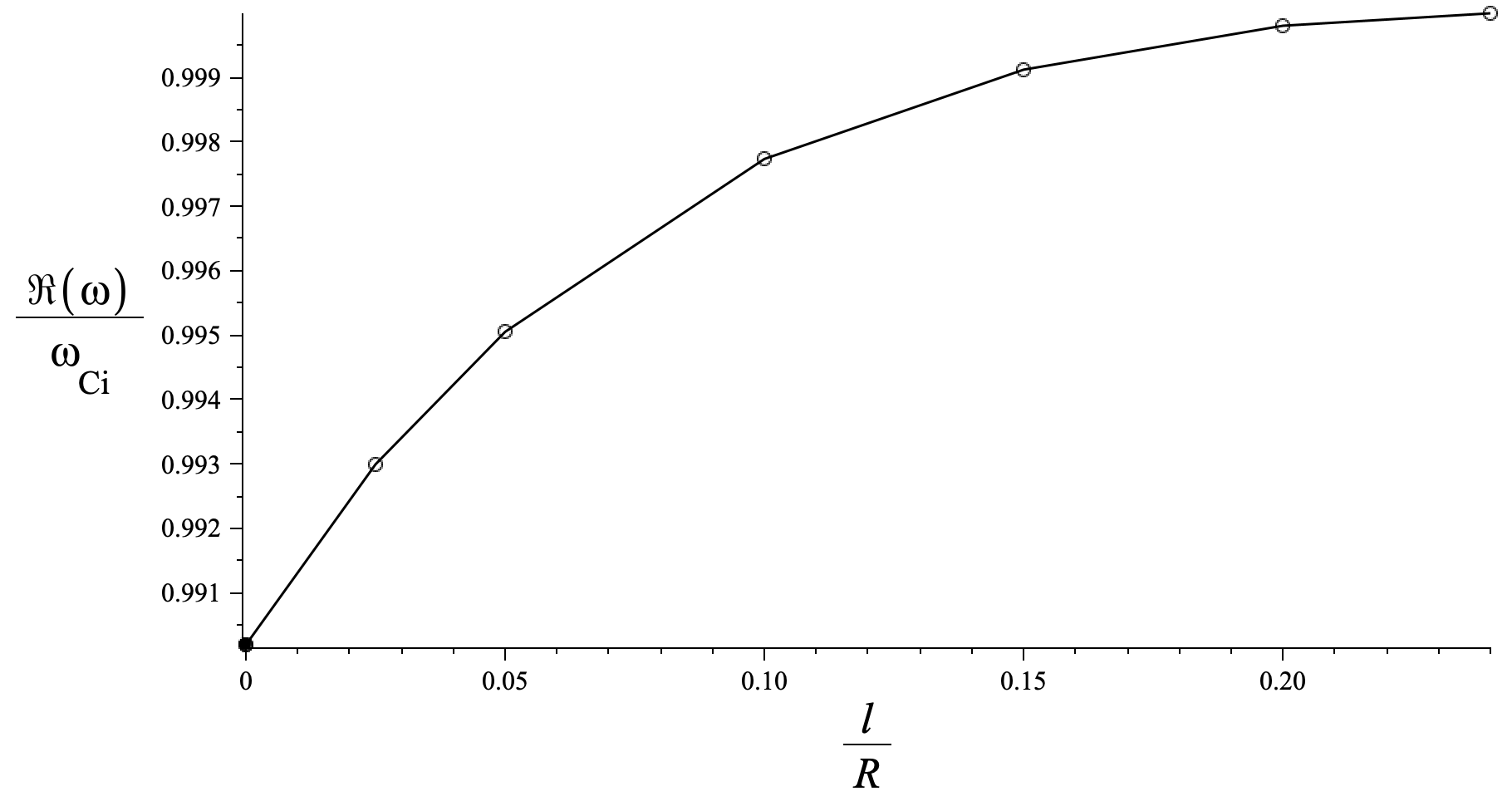}
   \caption{Real part of the frequency of the slow sausage quasimode (solid line) as a function of $l/R$ for $k_z R = 1$ (top), $k_z R = 3$ (middle), and $k_z R =5$ (bottom). The hollow circles represent the quasimode, with the frequency calculated from the dispersion relation obtained through the series method outlined in the previous sections. The full circle is the frequency of the slow surface sausage mode in the case of a discontinuous boundary, calculated from the dispersion relation in \citet{Edwin&Roberts1983} under the same physical conditions.}
              \label{SlowReal}%
    \end{figure}

\begin{figure}
   \centering
   \includegraphics[scale=0.28]{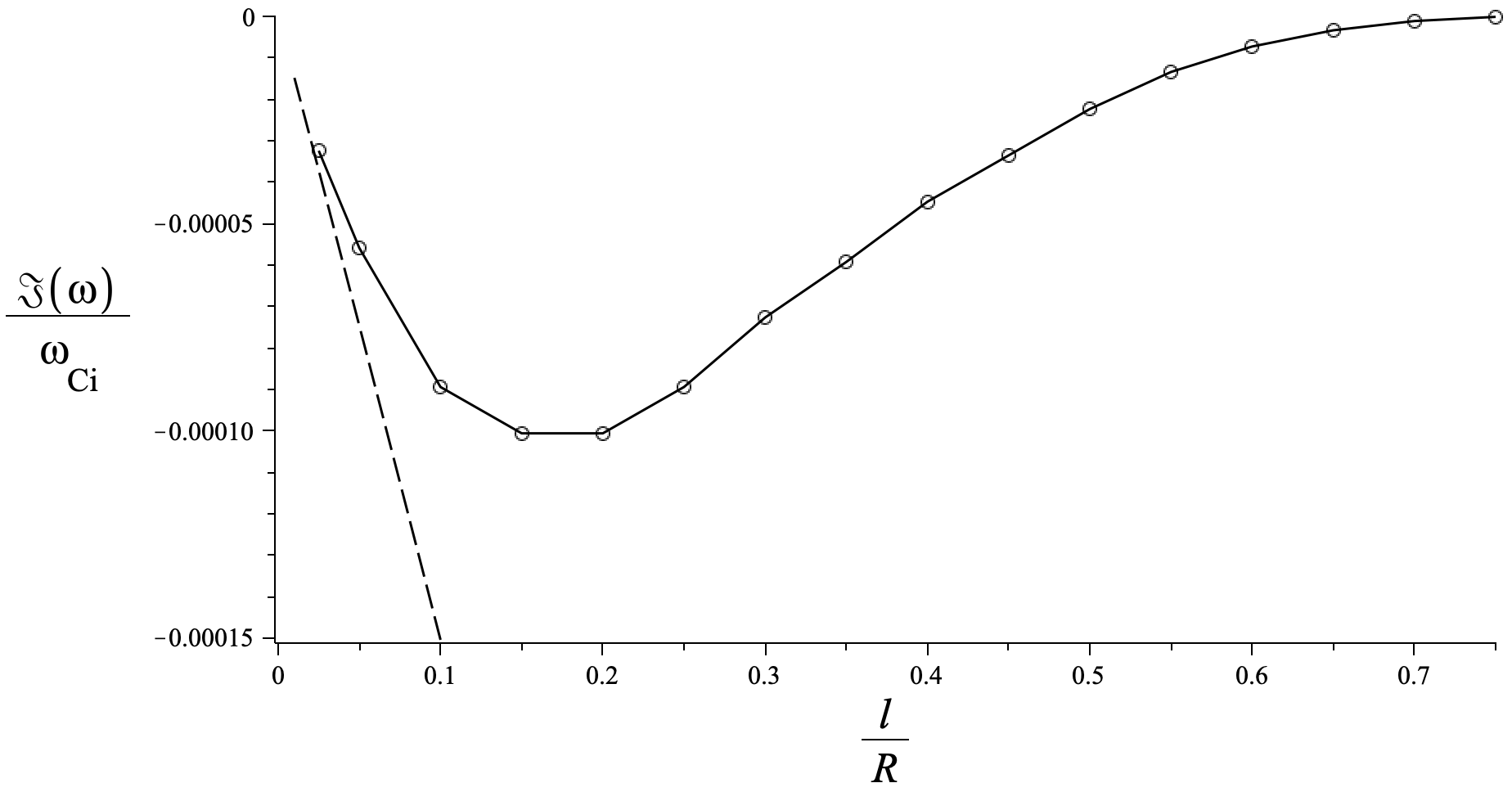}
   \includegraphics[scale=0.28]{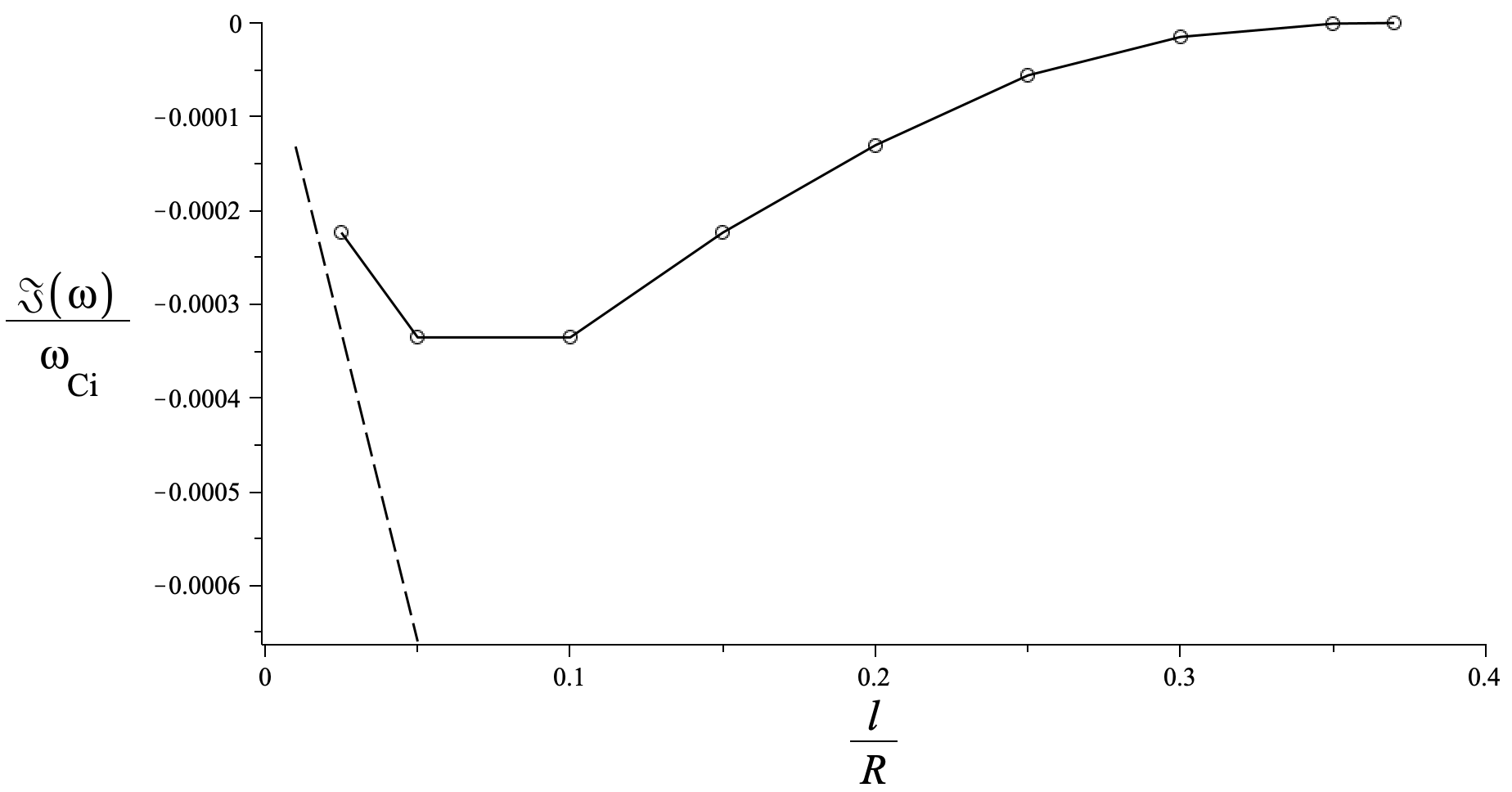}
   \includegraphics[scale=0.28]{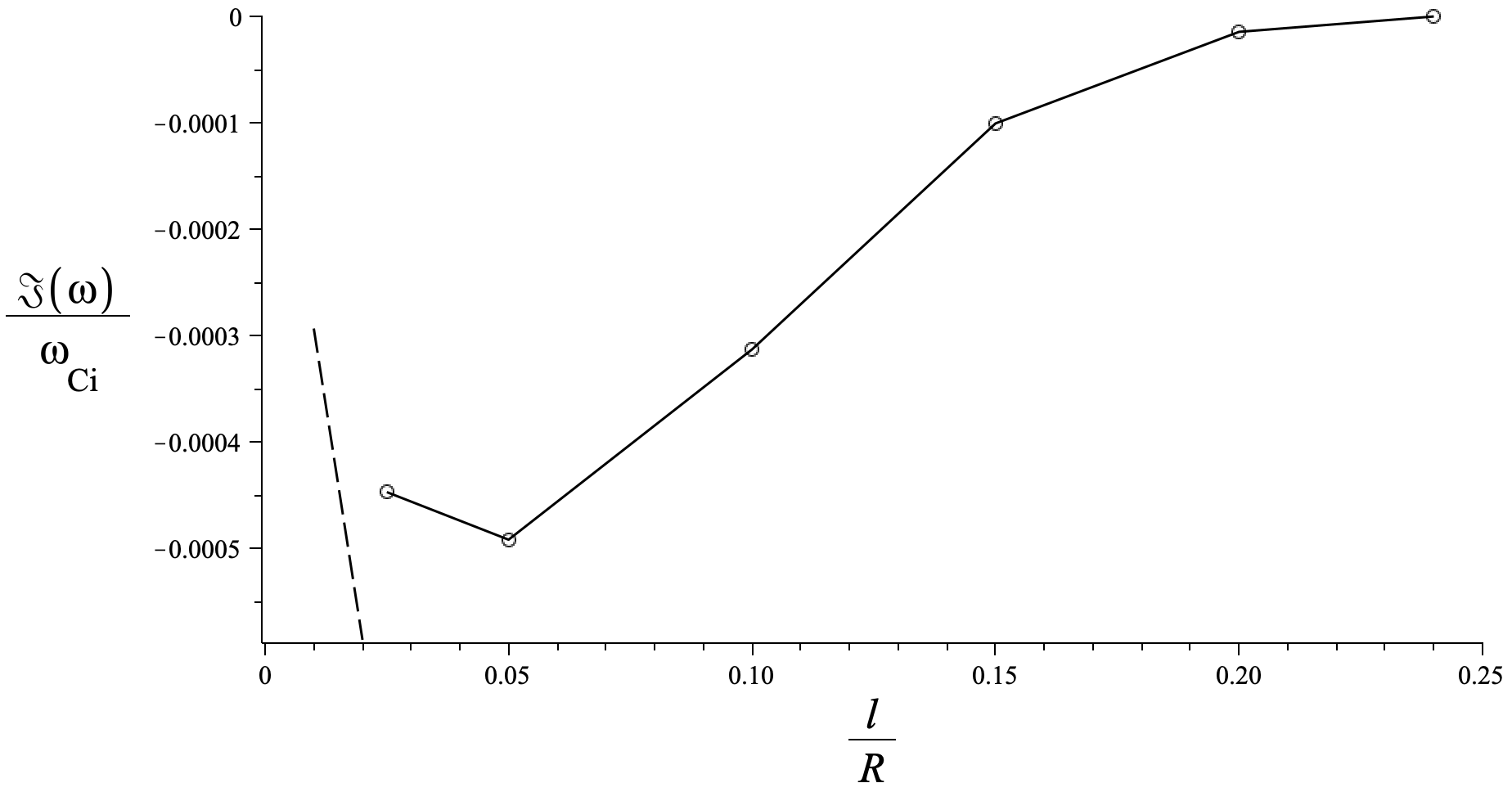}
   \caption{Imaginary part of the frequency of the slow sausage quasimode (solid line) as a function of $l/R$ for $k_z R = 1$ (top), $k_z R = 3$ (middle), and $k_z R =5$ (bottom) calculated from the dispersion relation obtained through the series method outlined in the previous sections. The modes represented here are the same as those from Fig. \ref{SlowReal}. The dashed line is the imaginary part of the frequency of the slow sausage quasimode calculated with the thin boundary approximation \eqref{DispTB}.}
              \label{SlowImag}%
    \end{figure}

For the sausage mode, the case is simplified as there is only one branch point $r_C$ in the complex $r$-plane. The quasimode frequencies $\o$ obtained with our series method, normalized with respect to the internal cusp frequency $\o_{Ci}$, are shown in function of $l/R$ in Fig. \ref{SlowReal} (real part) and Fig. \ref{SlowImag} (imaginary part) for $k_z R = 1$, $k_z R =3$ and $k_z R = 5$. These values for the longitudinal wavenumber correspond to those observed in slow sausage modes in photoshperic pores by \citet{GrantEtAl2015}, for which $k_z \in [1 , 5]$. Fig. \ref{SlowImag} also contains the imaginary part of the quasimode frequency calculated with the dispersion relation in the thin boundary approximation \eqref{DispTB}.

In Fig. \ref{SlowReal}, we notice that the quasimode frequency approaches the internal cusp frequency in its real part as $l/R$ is increased. This can also be seen in Figure 4 in \citet{ChenEtAl2018}, which is produced from numerical computations in the resistive MHD model and which shows the real part of the resistive eigenfrequency of the slow sausage mode as a function of $l/R$ in the same physical setup although with slightly different transition profiles. This is not taken into account in the analytical thin boundary (TB) approximation, where the oscillatory part of the frequency is assumed constant \citep{YuEtAl2017b}.

The imaginary parts of the quasimodes, in Fig. \ref{SlowImag}, display a behavior which might be unexpected. Initially, as $l/R$ is slightly increased the imaginary part is increased in absolute value as well, corresponding to a stronger damping. However, after some critical value of $l/R$ which depends on $k_z R$ has been reached, the absolute value of the imaginary part of the quasimode becomes smaller again and stays on this decreasing path. This stands in contrast with both the thin tube thin boundary (TTTB) approximation for nonaxisymmetric modes and the TB approximation for the slow surface mode, as the damping is found to be monotonically increasing with $l/R$ in those cases. In Fig. \ref{SlowImag}, we can indeed see that the imaginary part of the frequency agrees very well with the result from the TB approximation when $l/R \to 0$, but that the two diverge from a certain value of $l/R$ onward as the TB estimation continues to increase indefinitely in absolute value.

The non-monotonic behavior in the imaginary part of the frequency was also found for fast kink quasimodes in the Alfv\'en continuum in the cold plasma case for certain transition profiles by \citet{SolerEtAl2013}, although for other profiles it is absent. It would thus seem the presence of this behavior depends on the transition profiles. \citet{Yu&VanDoorsselaere2019} also found that mode conversion in coronal loops is more efficient when the transition layer is thin, as long as the transition profile is approximately linear at the resonant position.

As $l/R$ is increased, the quasimode frequency comes very close to the internal cusp frequency $\o_{Ci}$. For values of $l/R$ above a certain critical value, which is about $0.75$ for $k_z R =1$, $0.40$ for $k_z R = 3$, and $0.25$ for $k_z R =1$, the discussed series method does not allow to properly compute the quasimode frequency anymore. The frequency $\o_{Ci}$ is a pole of the dispersion function as well as an accumulation point for the slow body modes that are not resonantly absorbed on the other side, making it impossible on a practical level to distinguish the quasimode from the pole and the infinite number of slow body modes in a small enough region around $\o_{Ci}$. It is hence not clear whether the frequency of the quasimode goes to the internal cusp frequency as a limit point when $l/R$ is increased, or if it goes past the internal cusp frequency and leaves the cusp continuum in its real part. It would then still be within the Alfv\'en continuum, which overlaps with part of the cusp continuum in the present photospheric conditions. However, as sausage modes are not resonantly absorbed in the Alfv\'en continuum in the present model \citep{SakuraiEtAl1991}, if a complex frequency were to be found outside the cusp continuum for these higher values of $l/R$ it should probably be discarded as only a mathematical continuation without physical sense anymore since there would be no continuum modes for the slow sausage mode to couple to.

\begin{figure}
   \centering
   \includegraphics[scale=0.33]{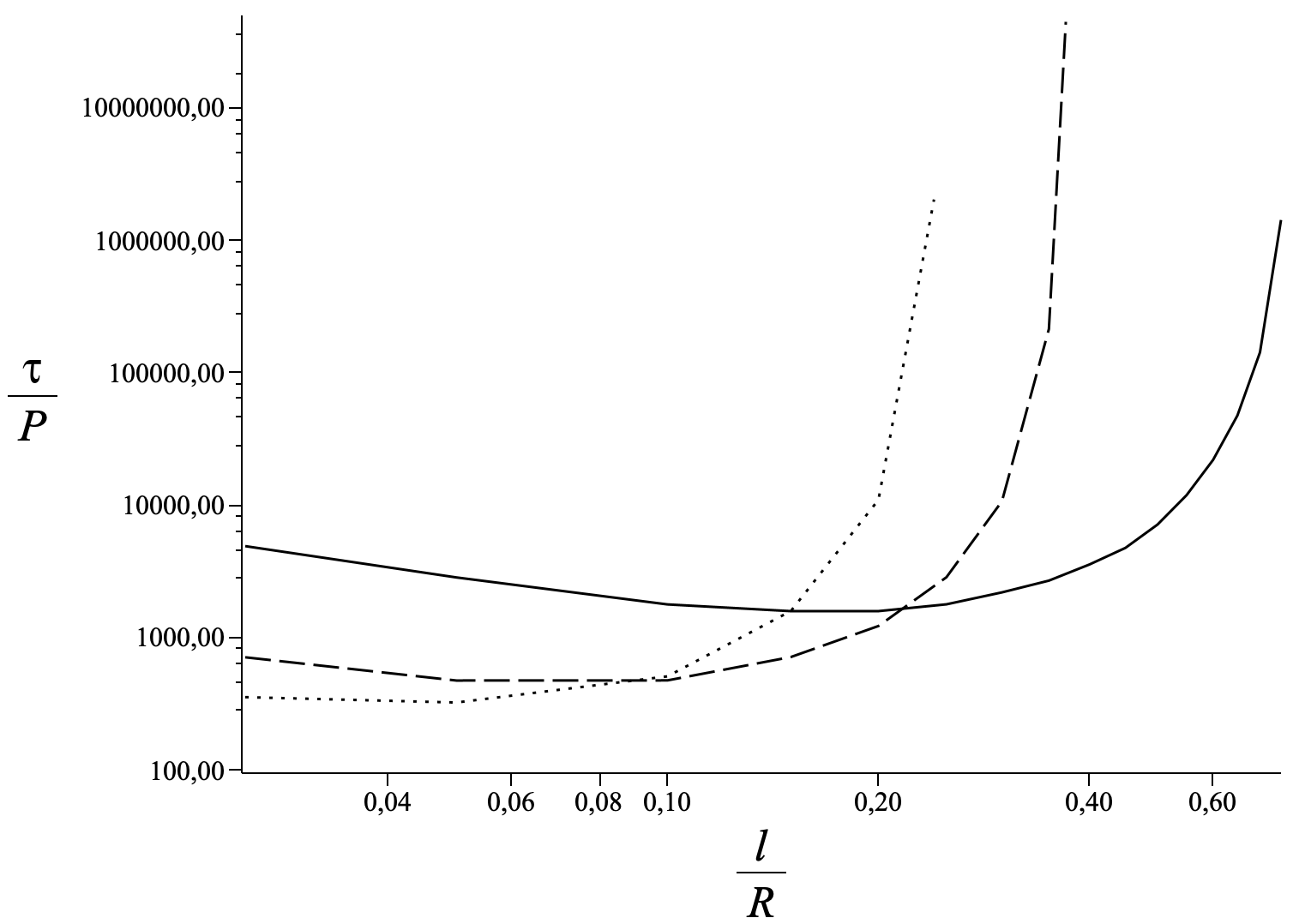}
      \caption{Damping time-to-period ratio $\tau/T$ as a function of $l/R$, for $k_z R = 1$ (solid line), $k_z R = 3$ (dashed line), and $k_z R = 5$ (dotted line). Both axes of the plot are in logarithmic scale.}
         \label{tauOverT}
   \end{figure}

We can also look at the ratio of the damping time $\tau = 1/\abs{\text{Im}[\o]}$ to the period $T = 2 \pi/\abs{\text{Re}[\o]}$ of the quasimode. This is shown in Fig. \ref{tauOverT}, where $\tau/T$ is plotted as a function of $l/R$ for the three different values of $k_z R$ considered before. The lowest value we find for $\tau/T$ is about $300$ at $l/R=0.05$ and $k_z R = 5$. This is larger than the damping time-to-period ratio of the resistive slow surface sausage eigenmode studied by \citet{ChenEtAl2018}, who found a $\tau/T \approx 100$ for $k_z R = 4.3$ and $\tau/T \approx 200$ for $k_z R = 2$, at $l/R=0.1$. This higher damping in the resistive case is to be expected of course, as resistivity provides an extra damping mechanism on top of resonant absorption. Their graph of $\tau/T$ also exhibits a convex shape, which can clearly be seen in Fig \ref{tauOverT} for $k_z R=1$ in our case too. As was also shown by \citet{ChenEtAl2018} through the TB dispersion relation obtained by \citet{YuEtAl2017b} and which we recovered in Eq. \eqref{DispTB} with our model, the analytical TB approximation does not yield this convex structure and it overestimates the damping as $l/R$ is increased. This also confirms our finding that, at least for some transition profiles in the inhomogeneous layer, the damping of the mode does not monotonically increase as the transition layer width $l/R$ increases. 

We see that the ideal results for slow sausage modes in the cusp continuum derived in this paper match the resistive results from \citet{ChenEtAl2018} quite well. This also confirms independently what \citet{ChenEtAl2018} concluded from their numerical study in resistive MHD, namely that the efficiency of resonant absorption of slow surface sausage modes in the cusp continuum in photospheric conditions is quite low and that other processes such as resistive damping might be more efficient for those modes.

\begin{figure*}
   \includegraphics[scale=0.25]{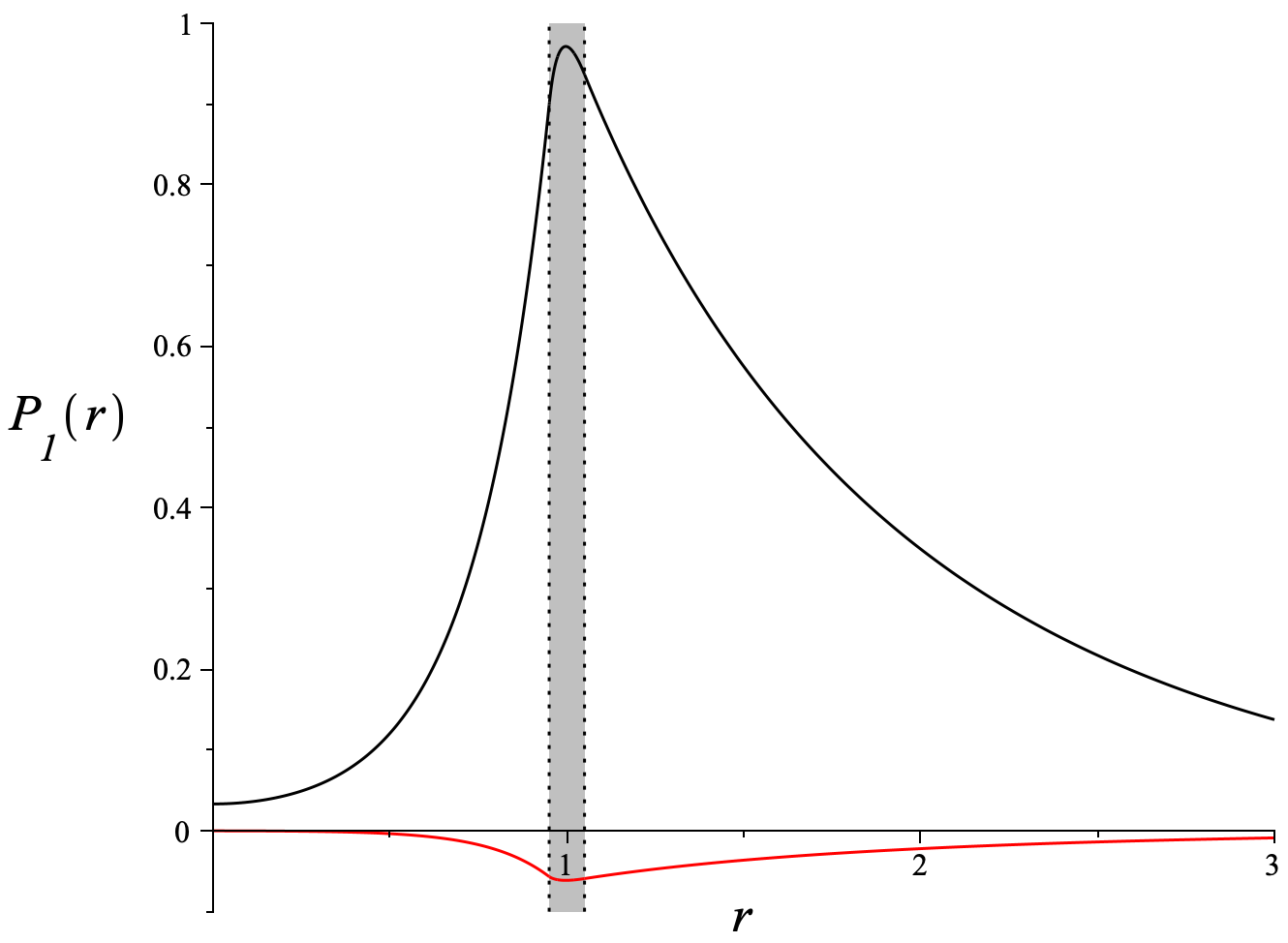}
   \includegraphics[scale=0.25]{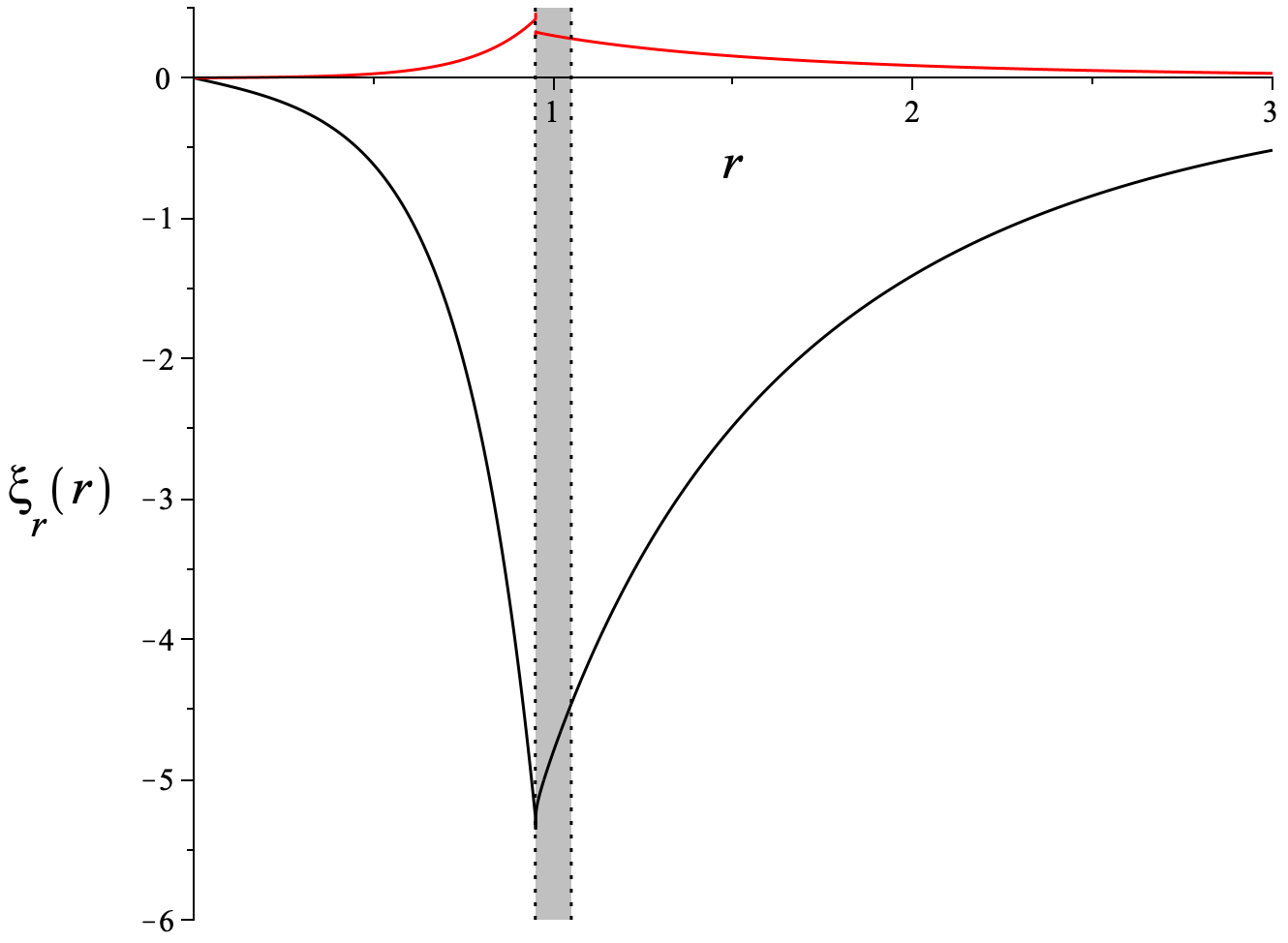}
   \includegraphics[scale=0.25]{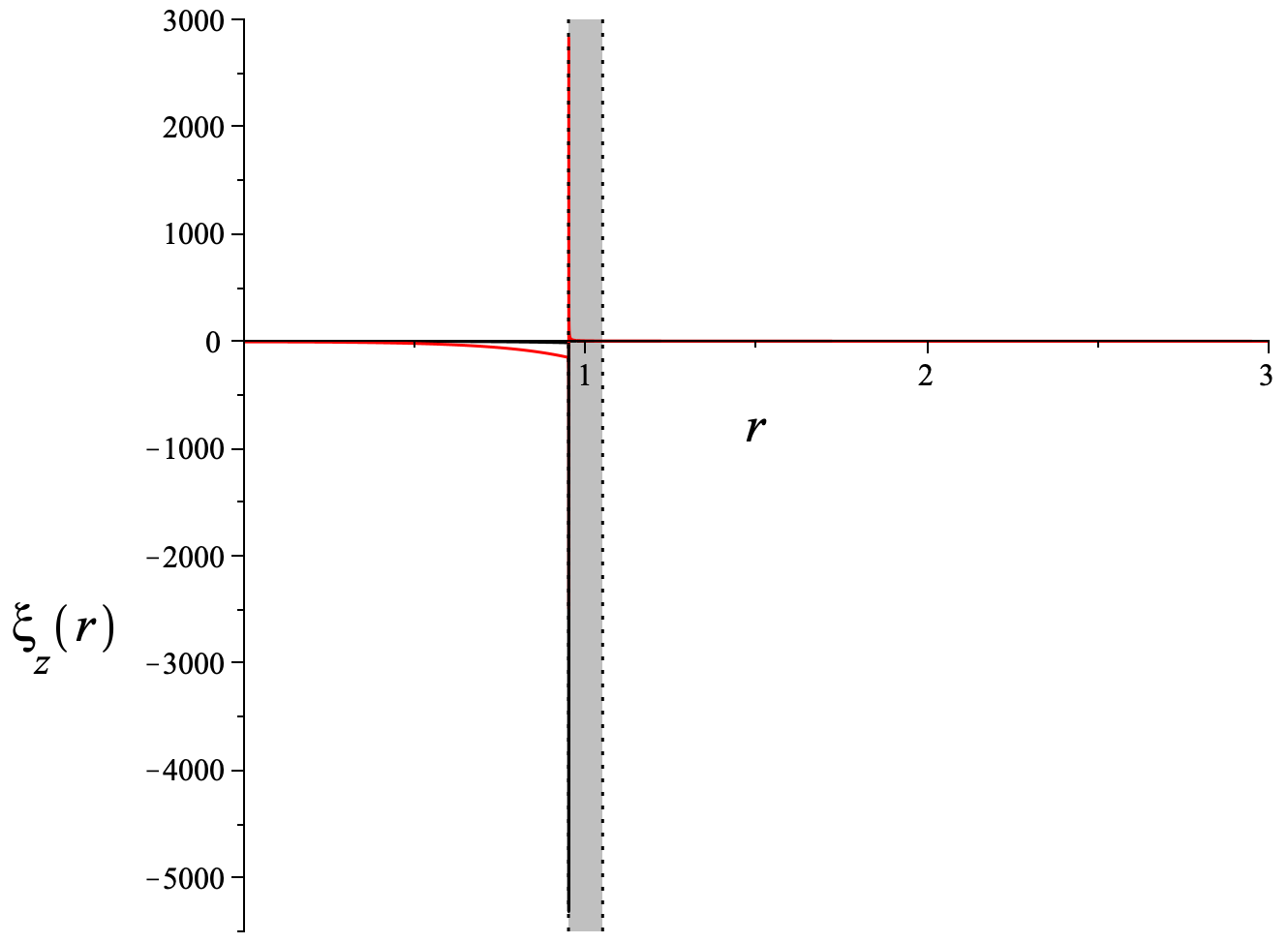}
   \caption{Real (black) and imaginary (red) parts of the quasimode perturbations $P_1$, $\xi_r$, and $\xi_z$, with $k_z R=1$ and $l/R=0.1$. The values of $r$ are normalized to $R$. The linear transition profile defined by Eqs. \eqref{v_C^2Linear}-\eqref{v_s^2hat} is taken in the inhomogeneous layer, which is represented in gray.}
              \label{QuasiModePertL01Kz1}%
    \end{figure*}

\begin{figure*}
   \includegraphics[scale=0.25]{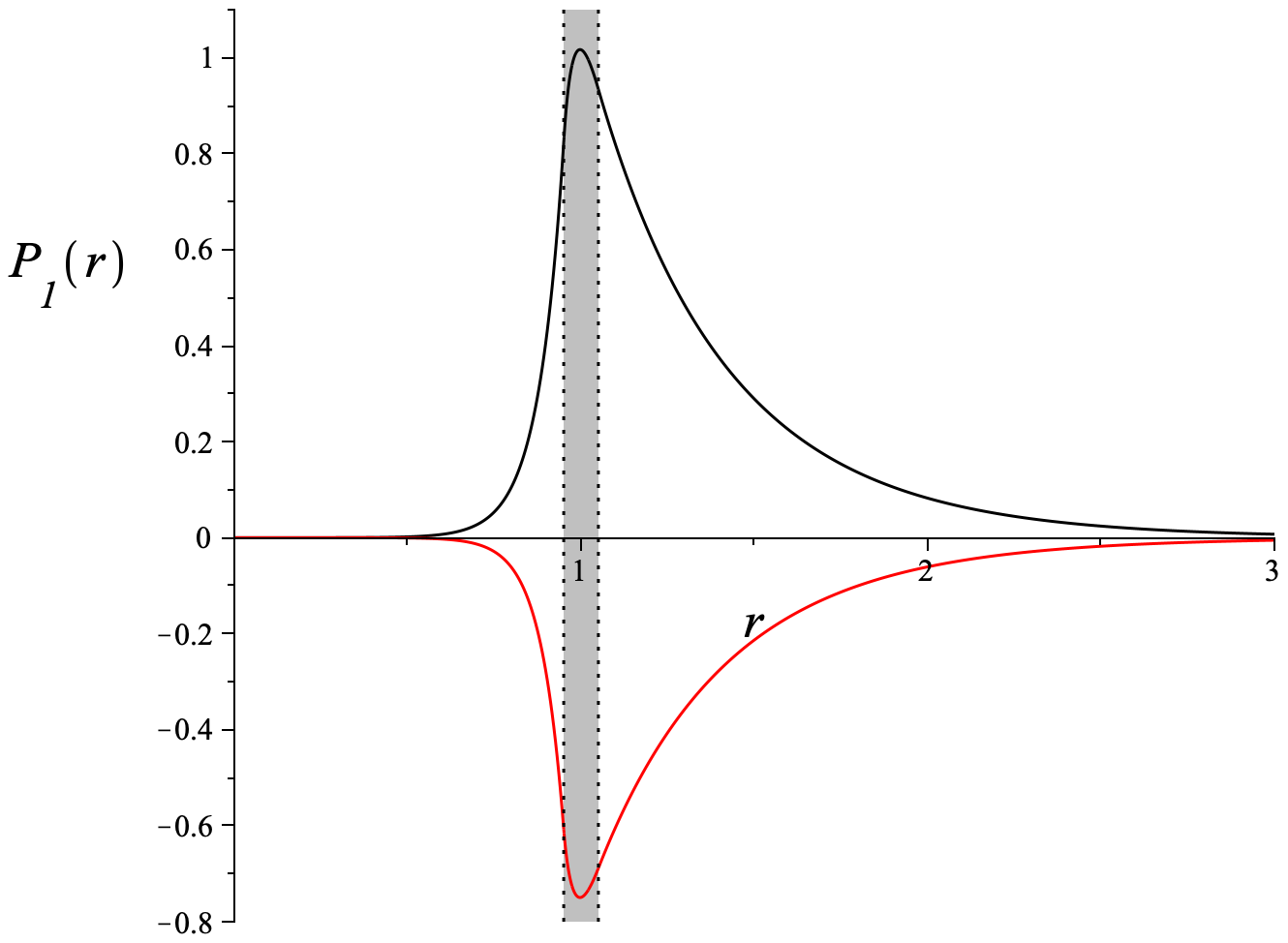}
   \includegraphics[scale=0.25]{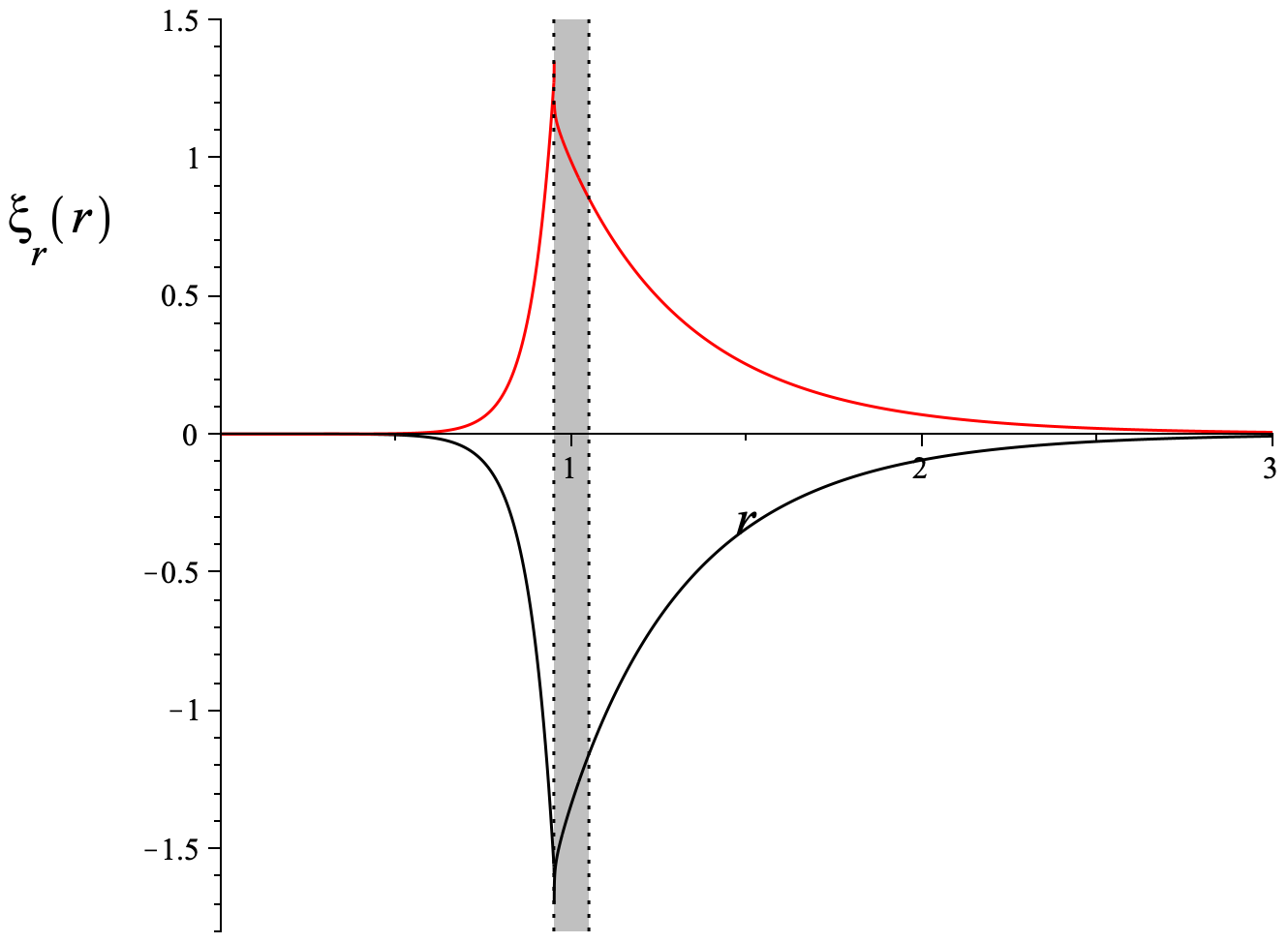}
   \includegraphics[scale=0.25]{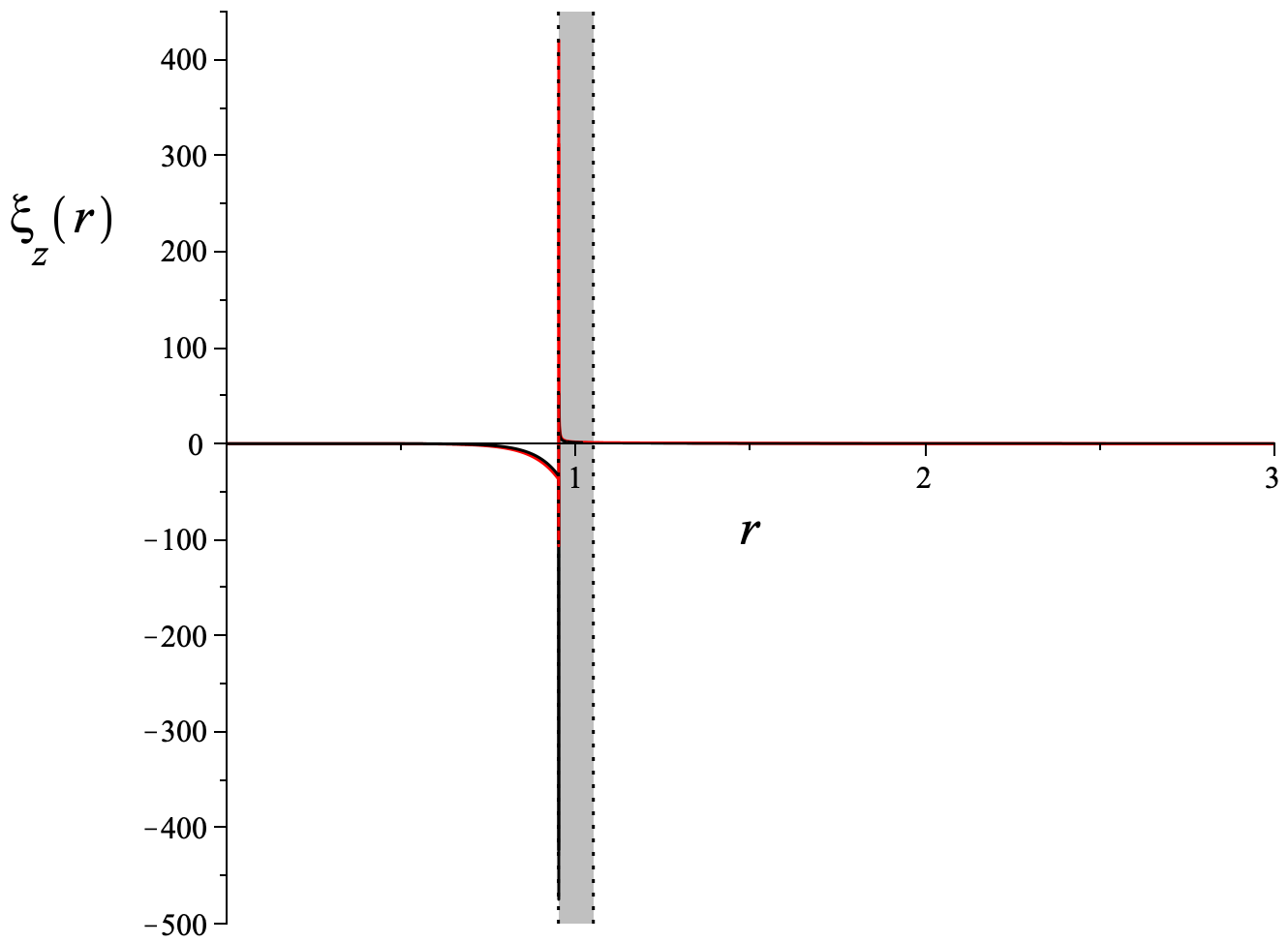}
   \caption{Real (black) and imaginary (red) parts of the quasimode perturbations $P_1$, $\xi_r$, and $\xi_z$, with $k_z R=3$ and $l/R=0.1$. The values of $r$ are normalized to $R$. The linear transition profile defined by Eqs. \eqref{v_C^2Linear}-\eqref{v_s^2hat} is taken in the inhomogeneous layer, which is represented in gray.}
              \label{QuasiModePertL01Kz3}%
    \end{figure*}

\begin{figure*}
   \includegraphics[scale=0.25]{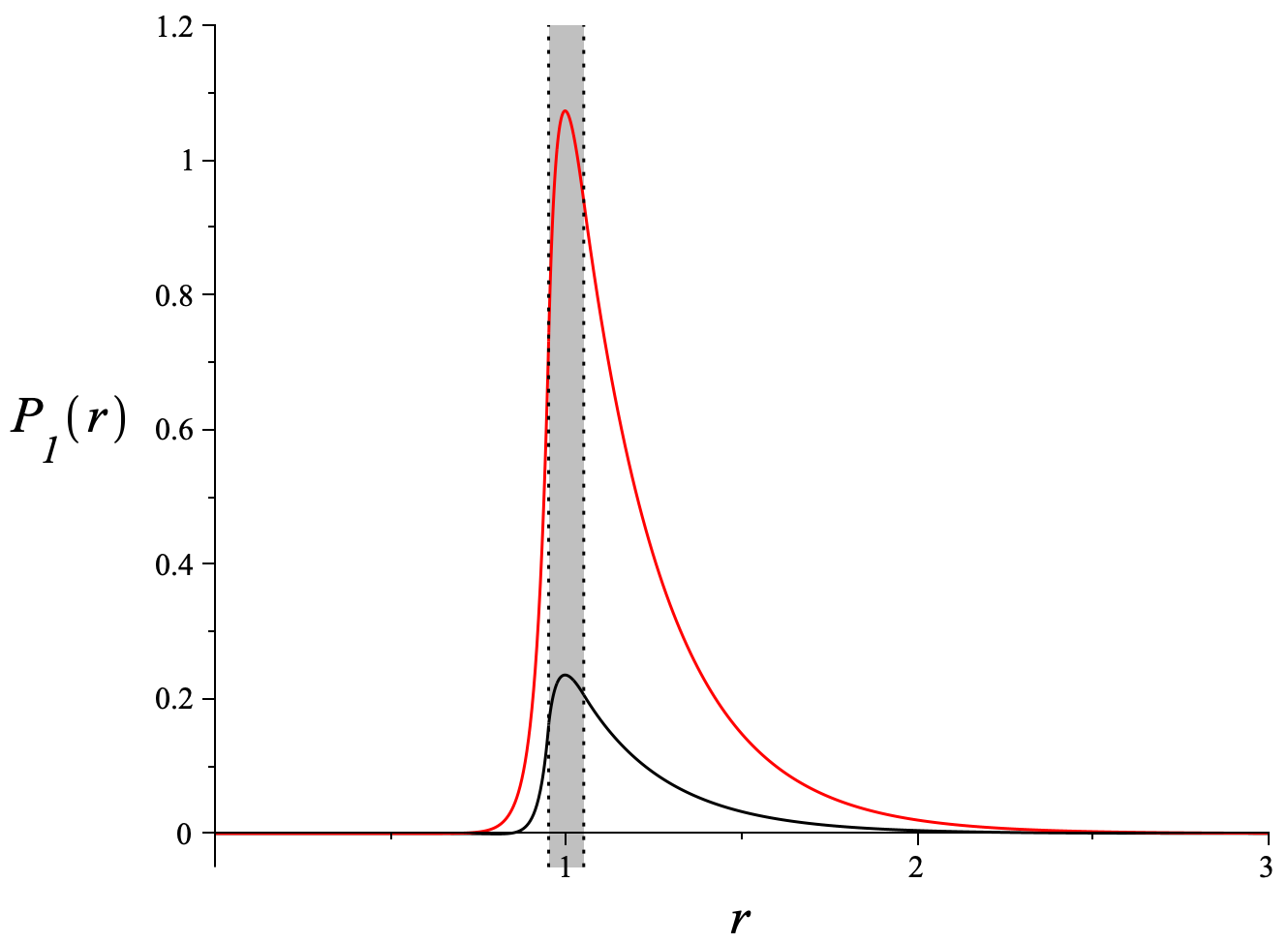}
   \includegraphics[scale=0.25]{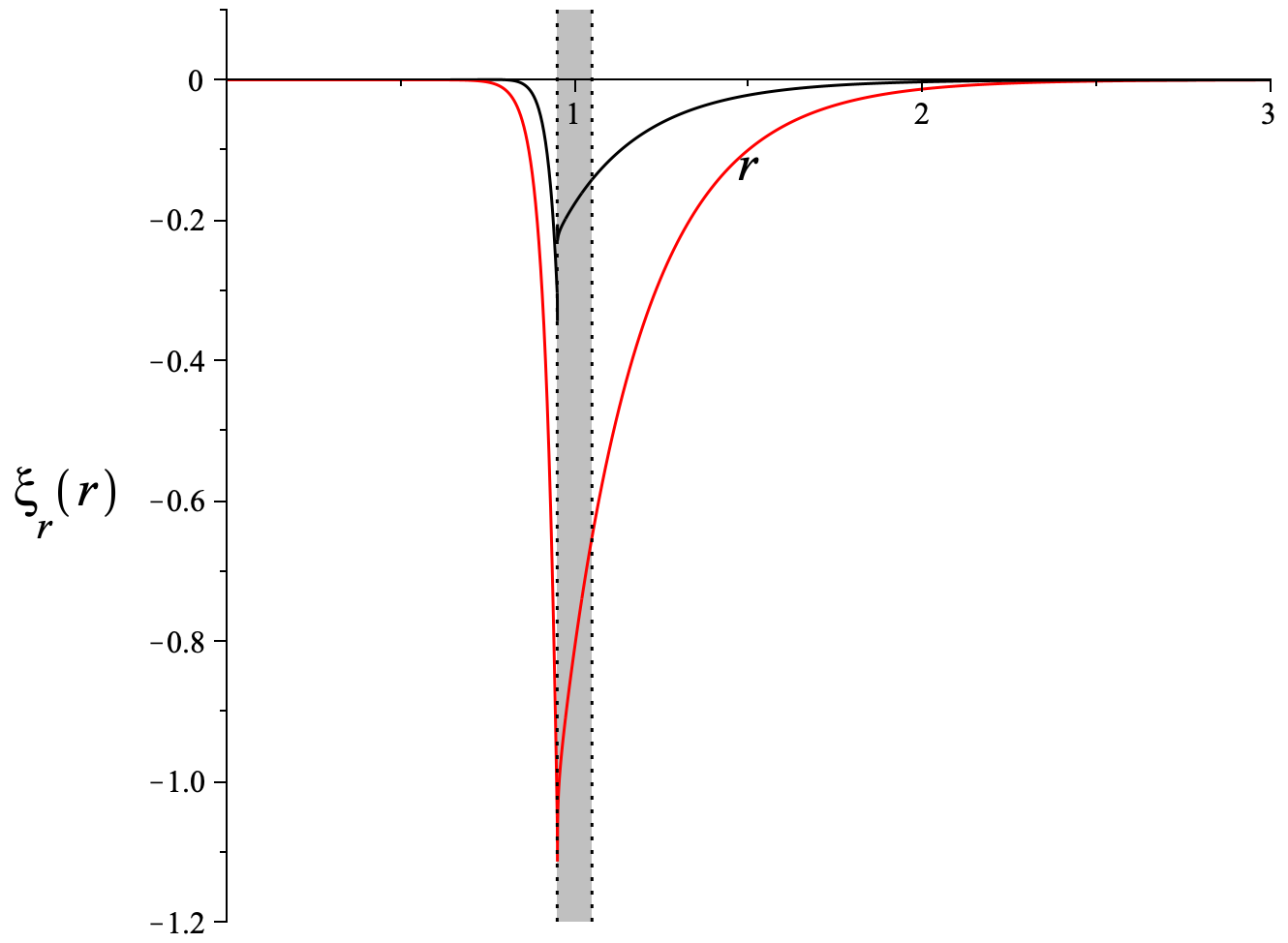}
   \includegraphics[scale=0.25]{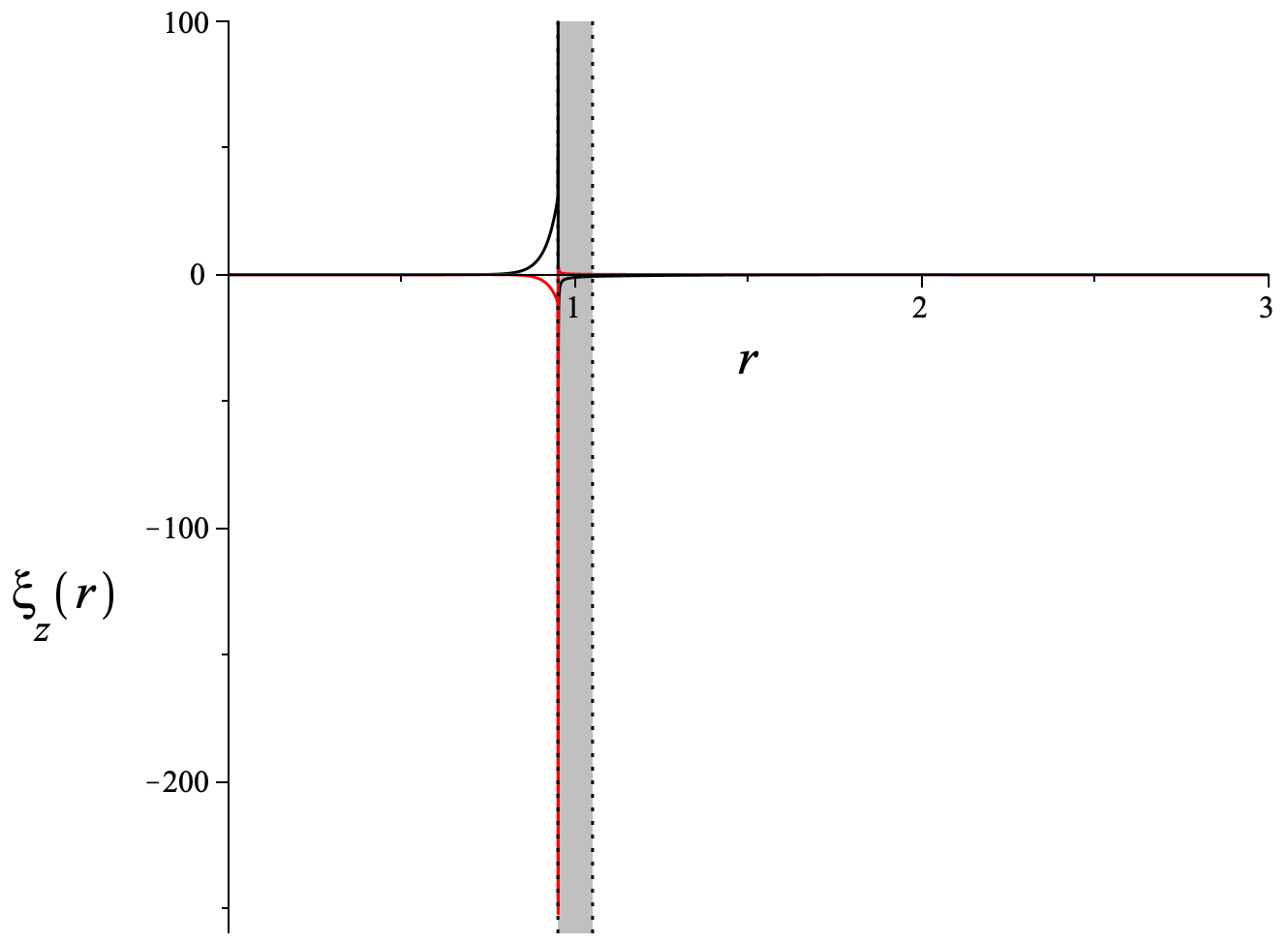}
   \caption{Real (black) and imaginary (red) parts of the quasimode perturbations $P_1$, $\xi_r$, and $\xi_z$, with $k_z R=5$ and $l/R=0.1$. The values of $r$ are normalized to $R$. The linear transition profile defined by Eqs. \eqref{v_C^2Linear}-\eqref{v_s^2hat} is taken in the inhomogeneous layer, which is represented in gray.}
              \label{QuasiModePertL01Kz5}%
    \end{figure*}

We can also look at the profiles of the perturbations of the quasimode. The profiles of $P_1$, $\xi_r$ and $\xi_z$ for a layer width of $l/R=0.1$ are shown in Fig. \ref{QuasiModePertL01Kz1} (for $k_zR=1$), Fig. \ref{QuasiModePertL01Kz3} (for $k_zR=3$), and Fig. \ref{QuasiModePertL01Kz5} (for $k_zR=5$). Since the singularity of Eq. \eqref{eqP1} lies in the complex plane, the solutions of this ODE are complex as well. This can be seen in the figures. We also notice that $\xi_z$ peaks at the resonant position near the left boundary of the layer, and is much larger than $\xi_r$ at the same position. This is to be expected, as \citet{SakuraiEtAl1991} showed that for modes which are resonantly absorbed in the cusp continuum the dominant contribution in the perturbations comes from $\xi_z$.

\begin{figure*}
   \includegraphics[scale=0.25]{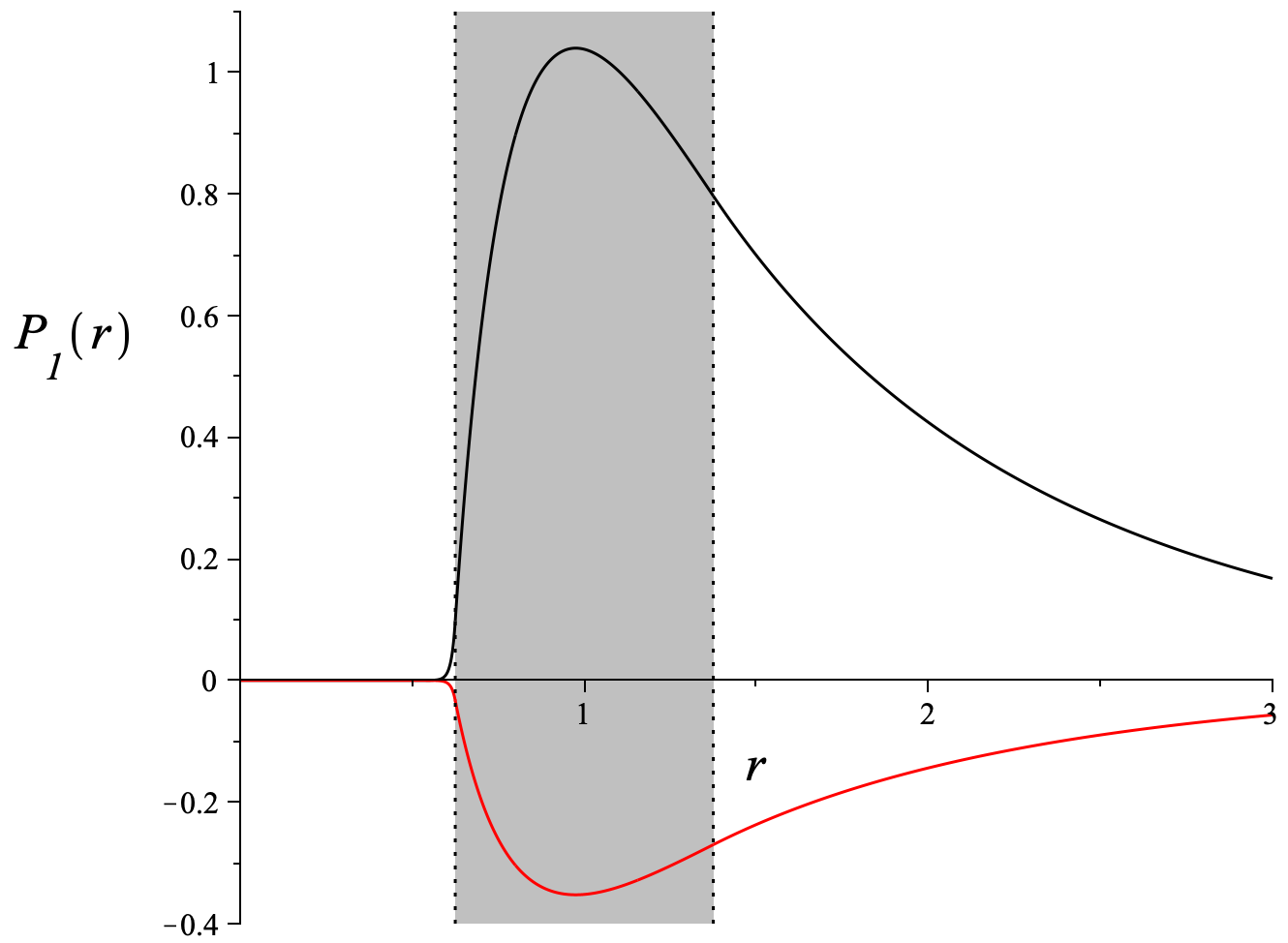}
   \includegraphics[scale=0.25]{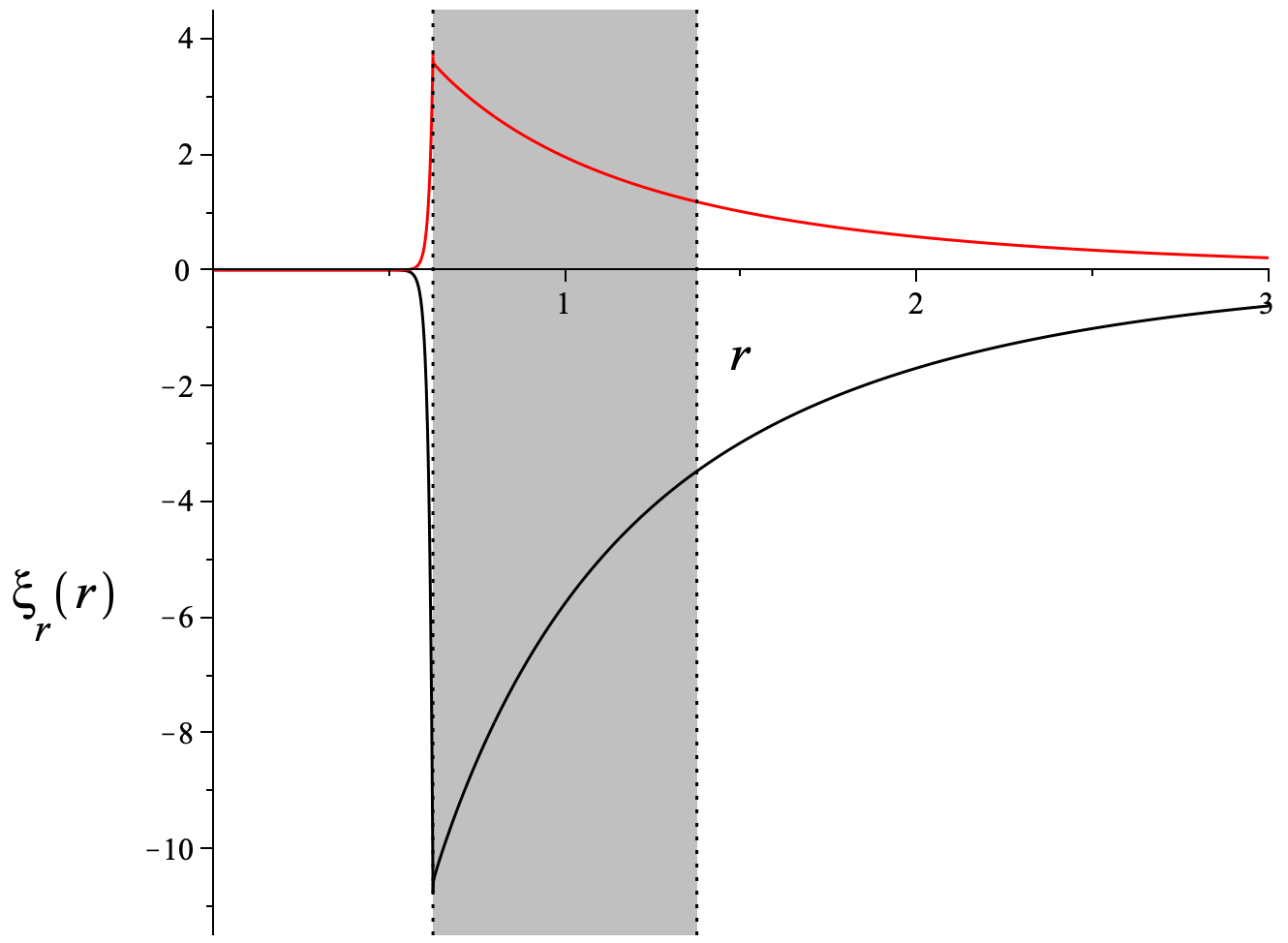}
   \includegraphics[scale=0.25]{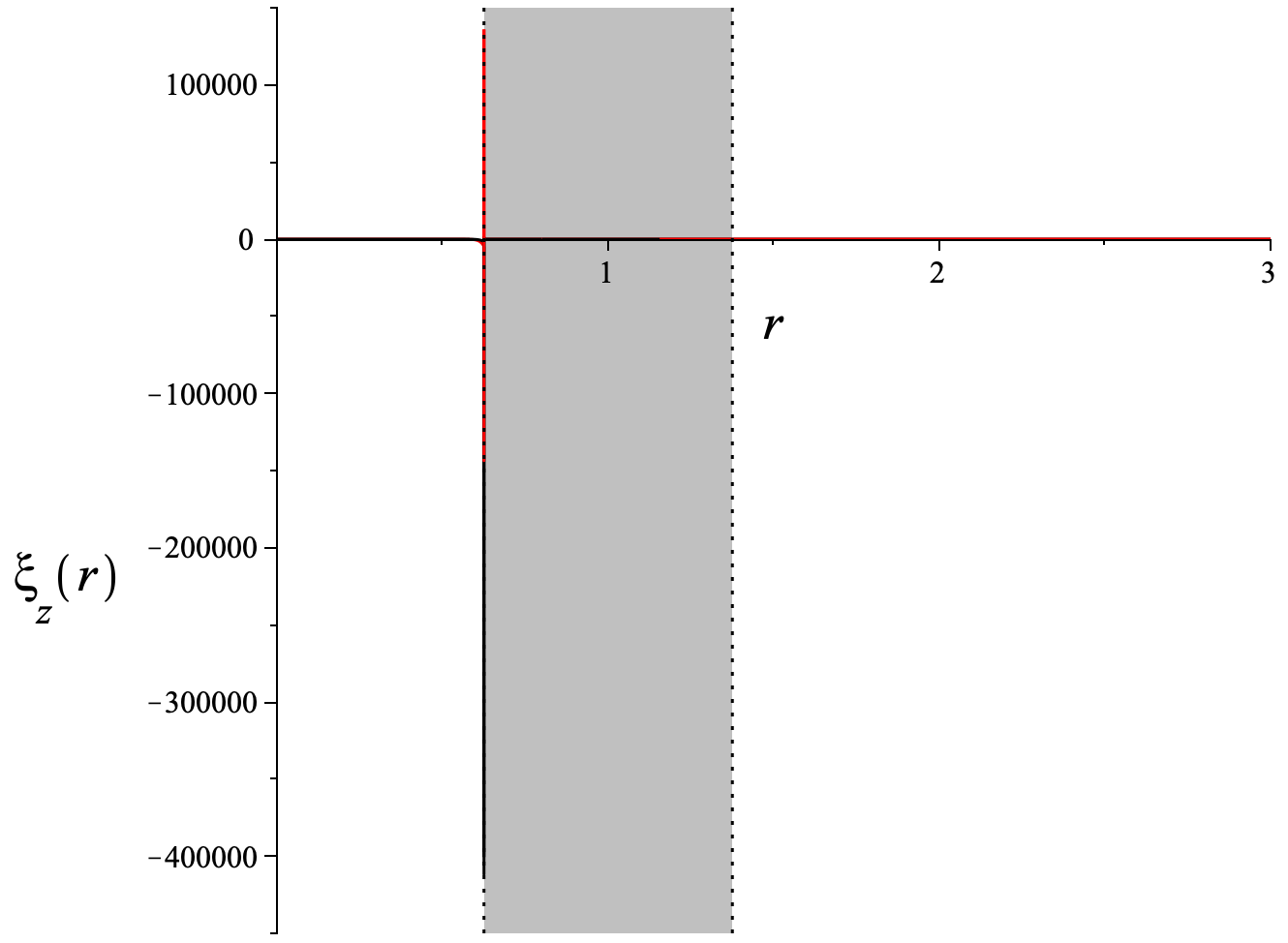}
   \caption{Real (black) and imaginary (red) parts of the quasimode perturbations $P_1$, $\xi_r$, and $\xi_z$, with $k_z R=1$ and $l/R=0.75$. The values of $r$ are normalized to $R$. The linear transition profile defined by Eqs. \eqref{v_C^2Linear}-\eqref{v_s^2hat} is taken in the inhomogeneous layer, which is represented in gray.}
              \label{QuasiModePertL075Kz1}%
    \end{figure*}

Figure \ref{QuasiModePertL075Kz1} shows the perturbations $P_1$, $\xi_r$ and $\xi_z$ for $l/R=0.75$ and $k_zR=1$. These are quite different in shape to the perturbations for $l/R=0.1$, as can be expected since the inhomogeneous layer is now much thicker. It was noted by \citet{SolerEtAl2013} that, since the singularity is not on the real axis, $P_1$ has a finite jump at the real part of the resonant position when plotting the profile over the positive real $r$-axis, even in the case of a thin layer. The jump conditions found in \citet{SakuraiEtAl1991}, which state that $P_1$ does not jump in a thin layer, are therefore only approximately true. However, \citet{SolerEtAl2013} found that the jump is much larger for a much thicker layer. As can be seen on Fig. \ref{QuasiModePertL075Kz1}, this is not the case here as the jump in $P_1$ remains very small and is invisible on the figure. The reason for this is that, unlike in the case discussed by \citet{SolerEtAl2013}, in our case the singularity stays close the real axis (and actually approaches it) when $l/R$ is increased from $0.1$ to $0.75$. This can be seen on Fig. \ref{SlowImag}.
    
We also note that the resonant position is very close to the inner boundary of the inhomogeneous layer. This is to be expected since, in the case of a discontinuous boundary without a layer, the slow surface mode has its frequency just below the internal cusp frequency in the interval $[\o_{Ce}, \o_{Ci}]$. As we saw in Fig. \ref{SlowReal}, the introduction of an inhomogeneous boundary layer changes the real part of the frequency of this mode only a bit. In contrast to this, the resonant position of the kink modes resonantly absorbed in the Alfv\'en continuum is situated near the middle of the inhomogeneous layer, as was shown by \citet{SolerEtAl2013}. We will recover the same results about the position of the two resonances for continuum modes in Section \ref{EigenfCont}.

\subsubsection{Kink and fluting modes}

For modes with $n \neq 0$, the overlapping of the cusp and Alfv\'en continua produces two branch points in the complex $r$-plane and thus a double branch cut on the overlapping part of the two continua (which is exactly the cusp continuum here) in the complex $\o$-plane. This renders the problem considerably more difficult.

Indeed, in the case of a single resonant point the Riemann sheets can be denoted with a so-called sheetnumber, depending on which branch of the logarithm is considered. The principal Riemann sheet then corresponds to a sheetnumber of $0$, whereas the neighboring sheets have sheetnumbers $1$ and $-1$. In the theory for handling quasimodes with the Laplace transform method as was laid out by both \citet{Sedlacek1971} and \citet{Goedbloed&Poedts2004}, the Bromwich contour is deformed in such a way that it surrounds the two zeros of the dispersion function which lie on the sheets with sheetnumber $1$ and $-1$ and have a negative imaginary part. These two zeros have an opposite oscillatory part of the frequency and an equal damping part, and represent a quasimode.

In the case of two resonant points, however, the introduction of the double branch cut on the cusp continuum implies that there are two dimensions in which a neighboring Riemann sheet can be considered. There are indeed two logarithms and hence in this case the sheets are denoted with a sheetcouple. The principal Riemann sheet is then denoted with the sheetcouple $(0,0)$, whereas there are eight neighboring Riemann sheets corresponding to the sheetcouples $(-1,1)$, $(0,1)$, $(1,1)$, $(-1,0)$, $(1,0)$, $(-1,-1)$, $(0,-1)$, and $(1,-1)$. We found that, in this case, there are zeros on various Riemann sheets but not on all of them. In addition, the zeros are not symmetric as in the $m=0$ case, in the sense that they neither have an opposite oscillatory part nor an equal imaginary part.
 
There would thus be several zeros with different values for $\o$ to consider when solving the Laplace transform in this situation, having to somehow deform the Bromwich contour to all eight neighboring Riemann sheets. It is not clear how the Laplace transform method has to be adapted to this much more intricate situation; we do not know if all zeros are physically relevant and represent quasimodes, or not. If only one would have to be considered, it is then unclear why the others would have to be rejected. This situation will thus not be handled here, but could be an interesting subject for a new paper.

\section{Eigenfunctions of continuum modes} \label{EigenfCont}

In the case of continuum modes, the frequency is real and the singularity lies on the real $r$-axis. As is explained for example in \citet{Goedbloed&Poedts2004}, the large Frobenius solution (i.e., the one containing the logarithm) is continuous across the singularity whereas the small Frobenius solution may jump. This implies that the general solution around a resonance point contains an additional arbitrary constant, which renders the system of equations formed by the boundary conditions inhomogeneous and thus allows them to always be fulfilled for continuum modes, independently of their frequency. Unlike for the quasimode, each basic independent solution of the ODE \eqref{eqP1} is real for the continuum mode, because the singularity now lies on the real axis instead of in the complex plane. With real coefficients in the general solutions, the continuum eigenfunctions are therefore either real or imaginary, but not complex.

\subsection{Sausage modes}

We recall that, under the photospheric conditions $(v_{Ae}, v_{si}, v_{se}) = (1/4,1/2,3/4) v_{Ai}$ assumed in the previous sections, the sausage mode is resonantly absorbed in the cusp continuum but not in the Alfv\'en continuum. The value of the additional arbitrary constant that arises from the resonant cusp position lying on the real $r$-axis then determines the amplitude of the perturbation.

In their Fig. 3, \citet{GoossensEtAl2021} plotted the modulus of the radial profile of some eigenfunctions of a resistive sausage eigenmode which corresponds to a continuum mode in ideal MHD, assuming a small resistivity and a sinusoidal transition profile in $v_C^2$ and $v_s^2$ in the inhomogeneous layer. We attempt to recover the eigenfunctions $\xi_r$ and $\xi_z$ from their figure by plotting the corresponding eigenfunctions of the continuum mode counterpart with our series method in ideal MHD. We assume the same resonant cusp position at about $r=0.955R$, the same values of $k_z R=2$ and $l/R=0.1$, but we have a linear transition profile for $v_C^2$ and $v_s^2$ instead of a sinusoidal one. 

\begin{figure}
   \centering
   \includegraphics[scale=0.25]{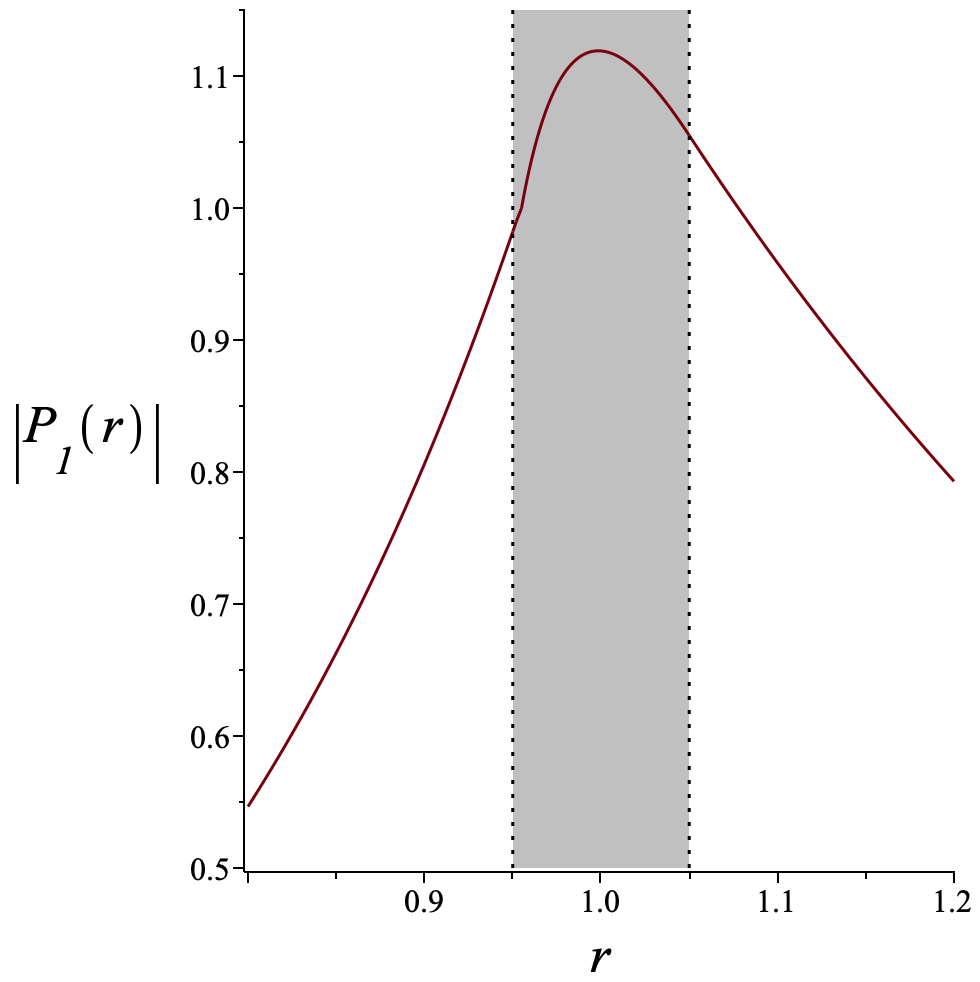}
   \includegraphics[scale=0.25]{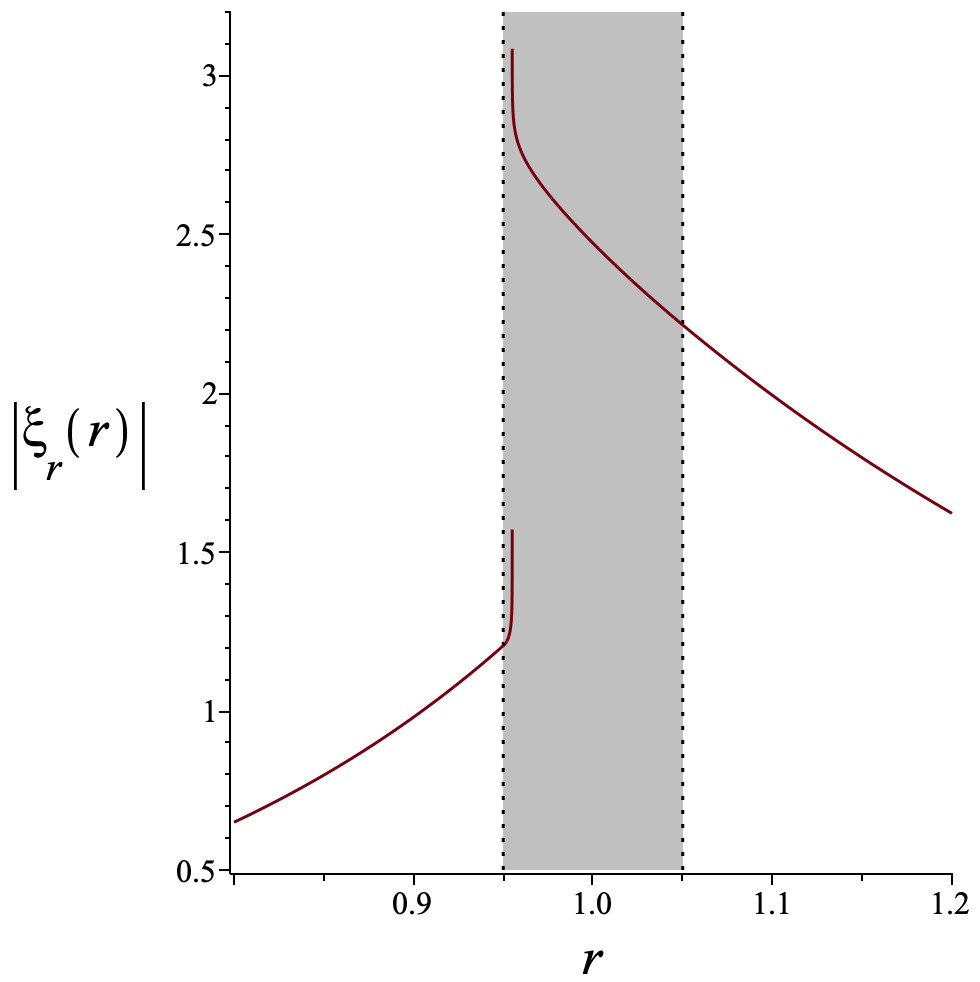}
   \includegraphics[scale=0.25]{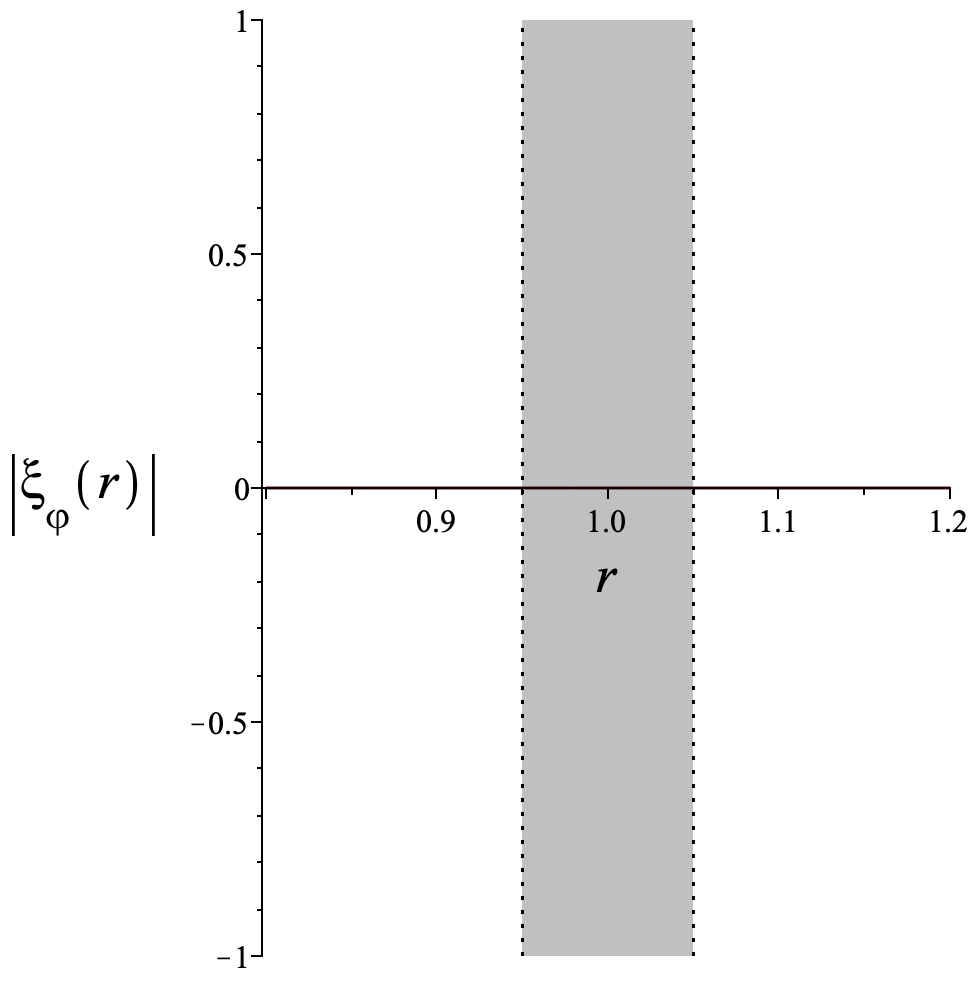}
   \includegraphics[scale=0.25]{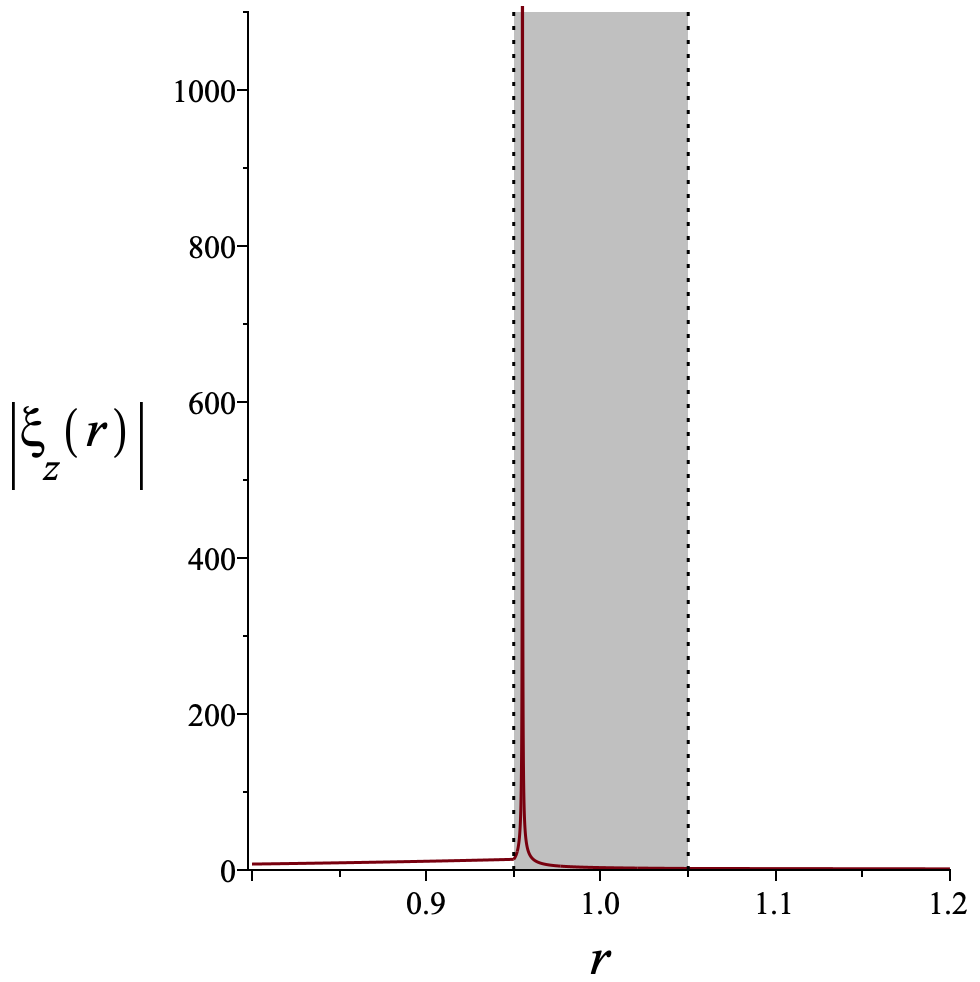}
   \caption{Moduli of the sausage continuum eigenfunctions $P_1$, $\xi_r$, $\xi_{\p}$ and $\xi_z$, with $r_C=0.955R$, $k_z R=2$ and $l/R=0.1$. The values of $r$ are normalized to $R$. The linear transition profile defined by Eqs. \eqref{v_C^2Linear}-\eqref{v_s^2hat} is taken in the inhomogeneous layer, which is represented in gray.}
              \label{eigenfM0}%
    \end{figure}

The obtained eigenfunction profiles for perturbed total pressure $P_1$, radial displacement $\xi_r$, azimuthal displacement $\xi_{\p}$, and longitudinal displacement $\xi_z$, are shown in Fig. \ref{eigenfM0}. The eigenfunctions $\xi_{\p}$ and $\xi_z$ can be computed from $P_1$ and are defined as follows:

\begin{align}
\xi_{\p} &= \d\frac{i n}{\rho r \l( \o^2 - \o_A^2 \r)} P_1 \text{,}\\
\xi_z &= \d\frac{i k_z v_s^2}{\rho \l( v_A^2 + v_s^2 \r) \l( \o^2 - \o_C^2 \r)} P_1 \text{.}\\
\end{align}

Our continuum eigenfunction for $\xi_z$ matches the resistive eigenfunction of \citet{GoossensEtAl2021} relatively well, but less so for $\xi_r$. In Fig. \ref{eigenfM0}, $P_1$ seems to have a sharp bend at $r_C$. However, from the expression of $P_1$ it can be verified that the graph is actually smooth and has a vertical tangent at that position, although it happens in a relatively small neighborhood around $r_C$. Consequently, the perturbation $\xi_r$, which is proportional to $\tod{P_1}{r}$ around $r_C$, seems to have a jump at this resonant position. It actually has a vertical asymptote there, as both sides go to infinity. The difference between the ideal continuum eigenfunctions and the resistive eigenfunctions might be due to resistivity smoothing over the apparent bend of $P_1$ at $r_C$.

The dominating perturbation at the cusp resonant position is $\xi_z$, in agreement with the analytical discussions by \citet{SakuraiEtAl1991} for example. Since this is a sausage mode, $\xi_{\p}$ is identically $0$.

\subsection{Kink modes}

In the case of kink modes, resonant absorption occurs both in the cusp and Alfv\'en continua. For each frequency in the overlapping continua, a continuum mode will now have two singularities and thus two additional arbitrary constants, say $D_C$ at the cusp resonant position and $D_A$ at the Alfv\'en resonant position. It is not entirely clear how these are related. As in the case of only one resonant position, the boundary conditions are fulfilled for every continuum frequency and the amplitude can be freely chosen. However, the inclusion of a second arbitrary constant entails that the radial profile itself differs in function of the ratio $D_C/D_A$.

\begin{figure}
   \centering
   \includegraphics[scale=0.25]{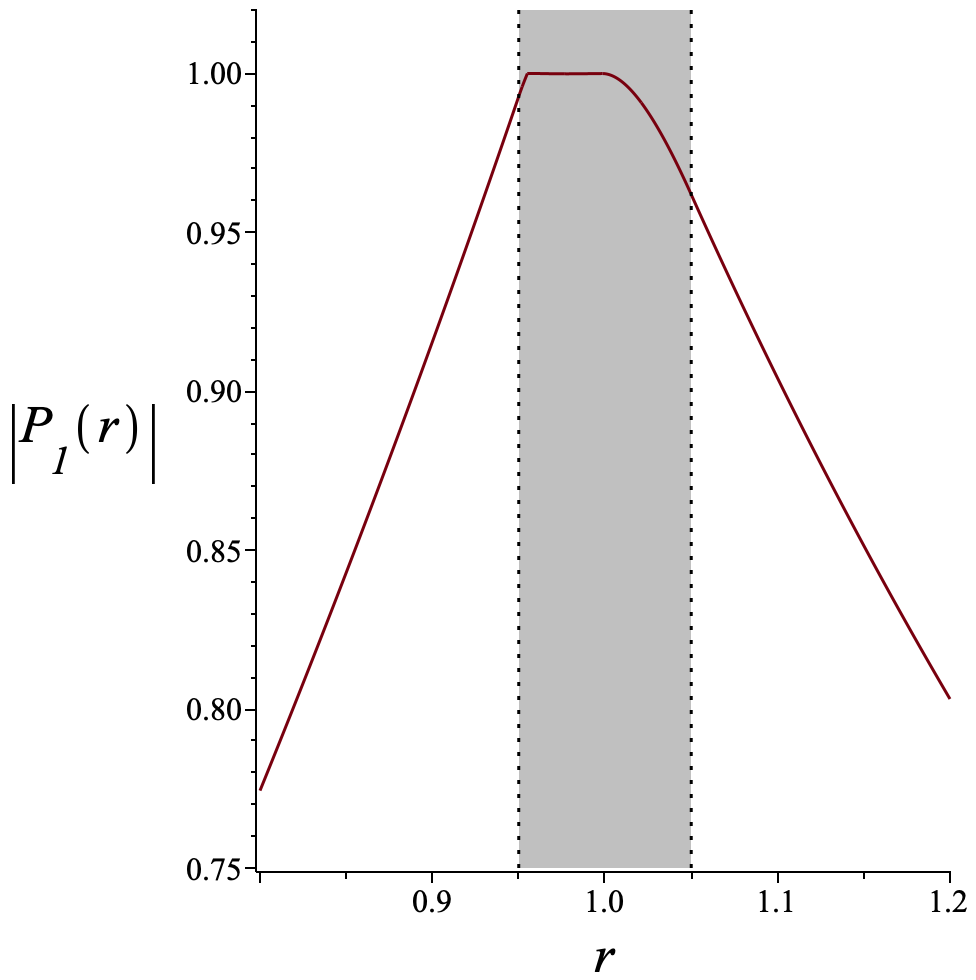}
   \includegraphics[scale=0.25]{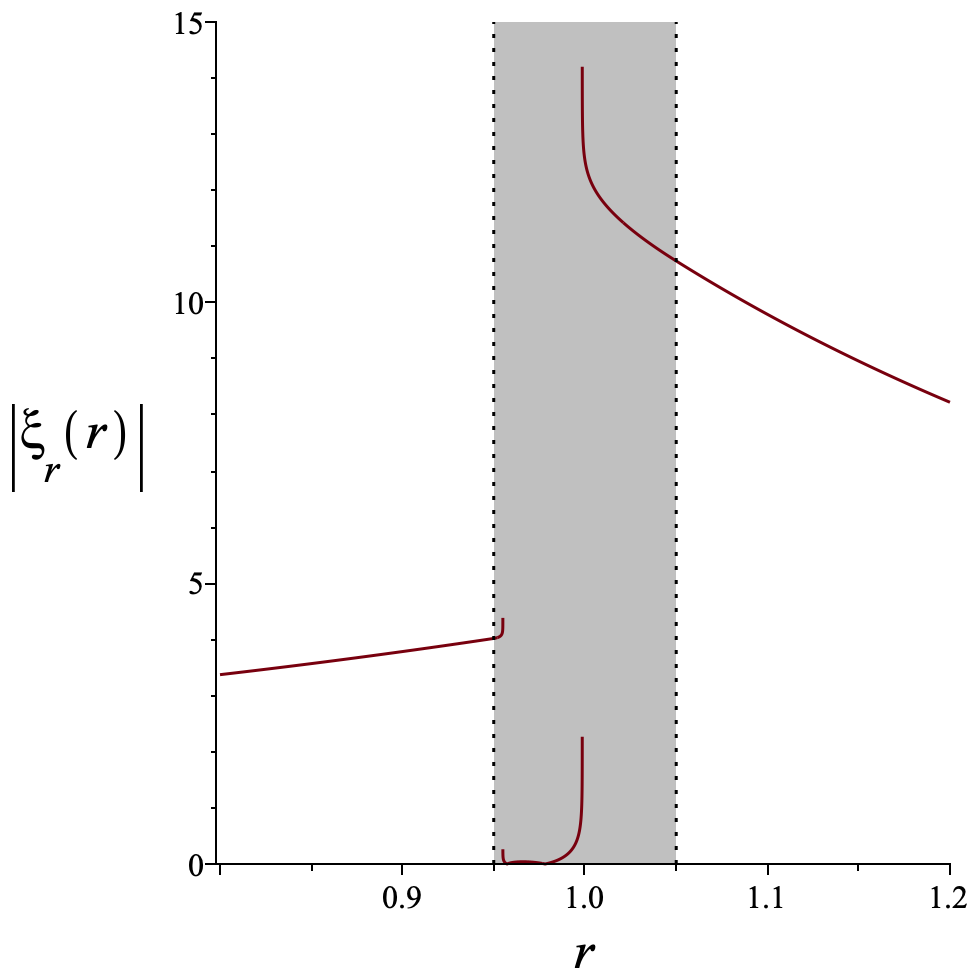}
   \includegraphics[scale=0.25]{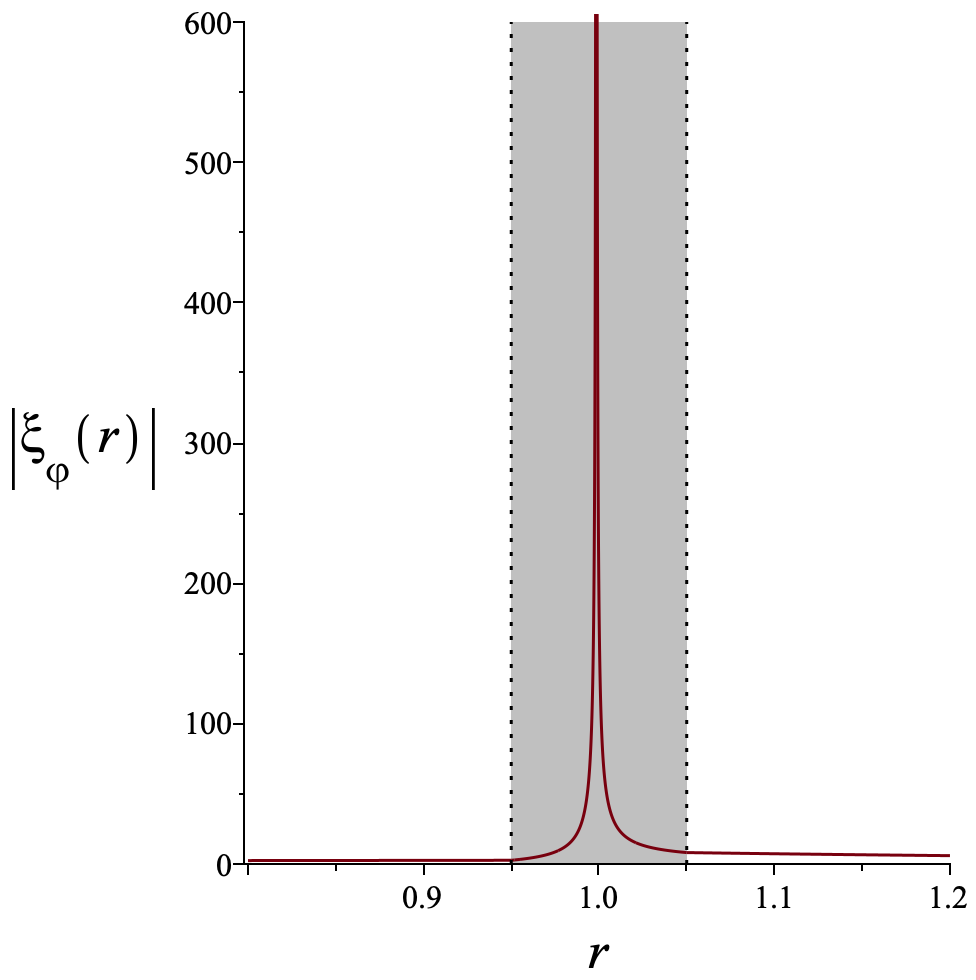}
   \includegraphics[scale=0.25]{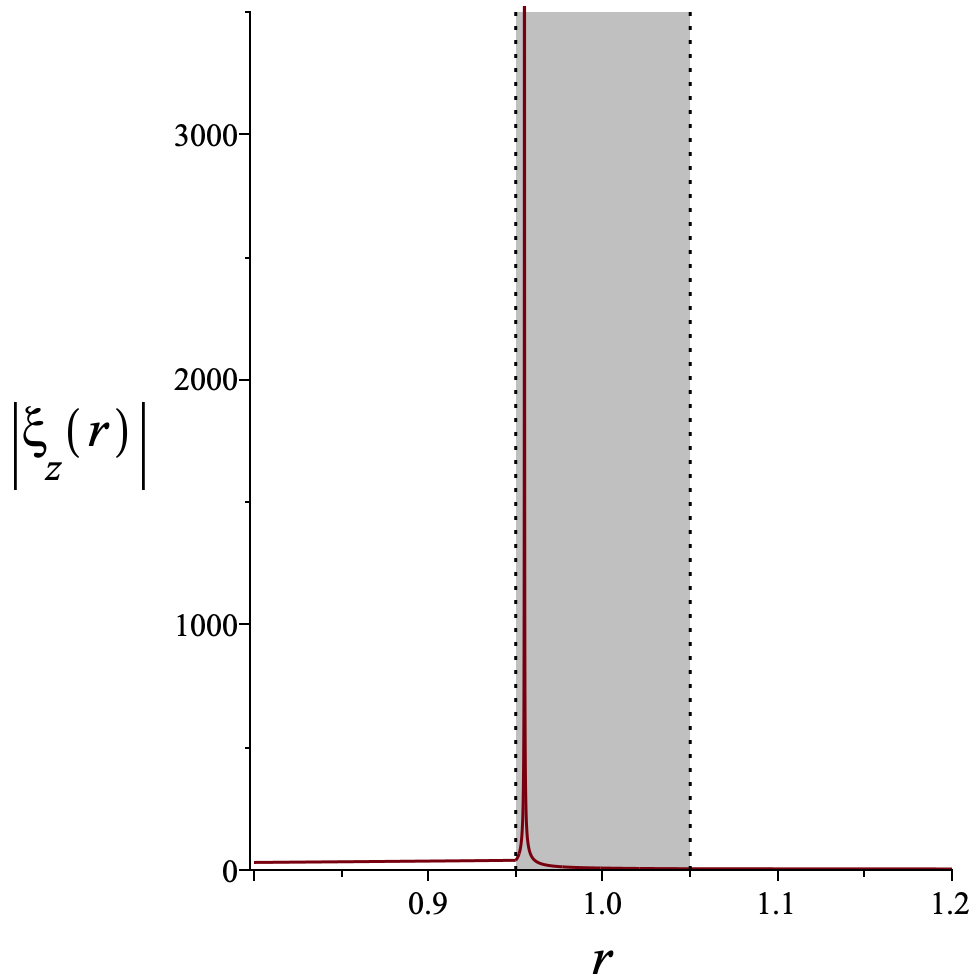}
   \caption{Moduli of the kink continuum eigenfunctions $P_1$, $\xi_r$, $\xi_{\p}$ and $\xi_z$, with $r_C=0.955R$, $k_z R=0.7$ and $l/R=0.1$. The values of $r$ are normalized to $R$. The linear transition profile defined by Eqs. \eqref{v_C^2Linear}-\eqref{v_s^2hat} is taken in the inhomogeneous layer, which is represented in gray. The ratio between the two arbitrary constants $D_C$ and $D_A$ has been taken equal to $1$ here.}
              \label{eigenfM1}%
    \end{figure}

The eigenfunctions $P_1$, $\xi_r$, $\xi_{\p}$ and $\xi_z$ of the continuum mode obtained from our series method in ideal MHD and corresponding to the resistive kink eigenmode shown by \citet{GoossensEtAl2021} in their Fig. 1 and Fig. 2 , are shown in Fig. \ref{eigenfM1}. To compare our eigenfunctions with theirs, we assume the same resonant cusp position at about $r=0.955R$, the same values of $k_z R=0.7$ and $l/R=0.1$, but we have again a linear transition profile for $v_C^2$ and $v_s^2$ instead of a sinusoidal one. Here, we chose a ratio of $D_C/D_A=1$, though a priori it seems any ratio would be acceptable as the boundary conditions are always fulfilled.

Our continuum eigenfunctions match the resistive eigenfunctions of Fig 1. in \citet{GoossensEtAl2021} quite well, except for $\xi_r$. In Fig. \ref{eigenfM1}, $P_1$ appears to have a sharp bend at $r_C$ again, but looking at its analytical expression we see that its radial profile is actually smooth and the graph has a vertical tangent at that position. The difference with the resistive eigenfunctions might again be due to resistivity smoothing over regions where the ideal profile is sharper.

Around the Alv\'en resonant position $r_A \approx R$, $P_1$ is visibly smooth on the plot, as can also be verified analytically. Indeed, $\xi_r$ is not directly proportional to $\tod{P_1}{r}$ at this position because of the extra factor $1/(\o^2-\o_A^2)$ in its definition. The eigenfunction $\xi_r$ has a vertical asymptote at $r_A$, whereas we have $\tod{P_1}{r}=0$ there. 

Furthermore, the dominating perturbation is $\xi_z$ at the cusp resonant position and $\xi_{\p}$ at the Alfv\'en resonant position. This is again in agreement with the analytical derivations of \citet{SakuraiEtAl1991}.

\begin{figure}
   \centering
   \includegraphics[scale=0.24]{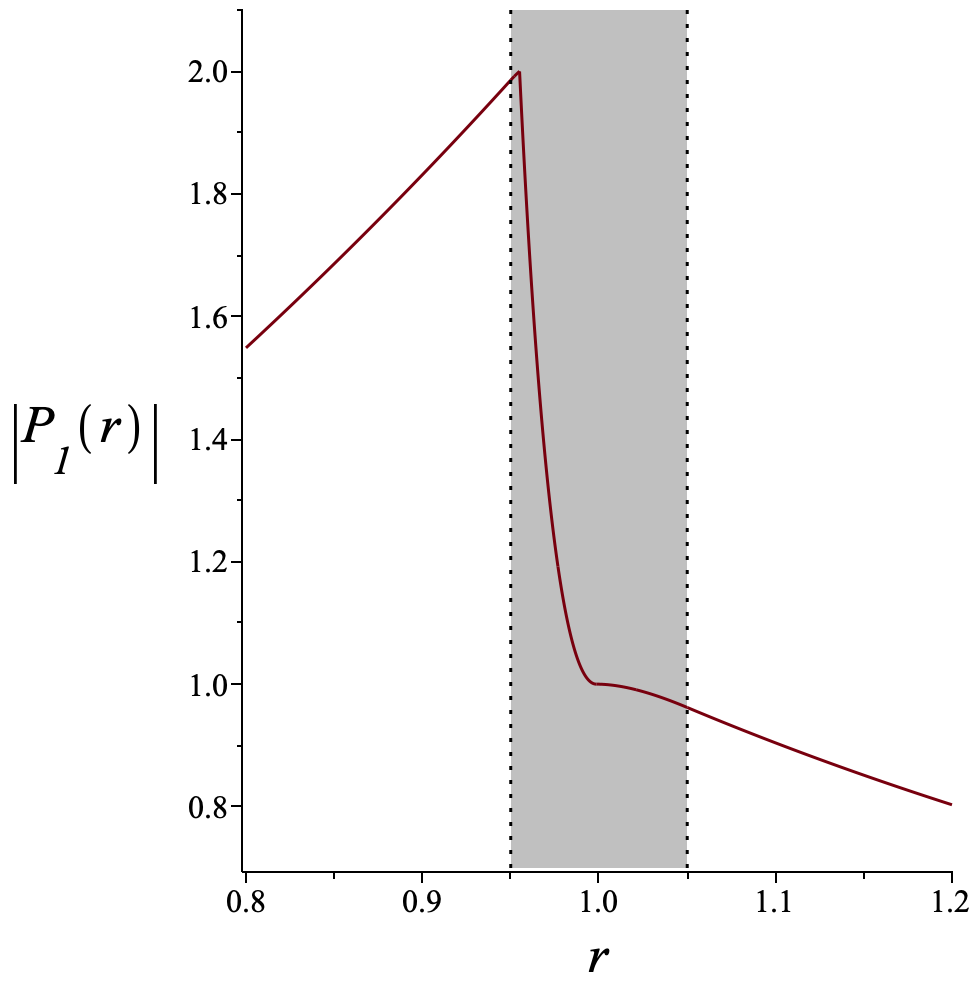}
   \includegraphics[scale=0.24]{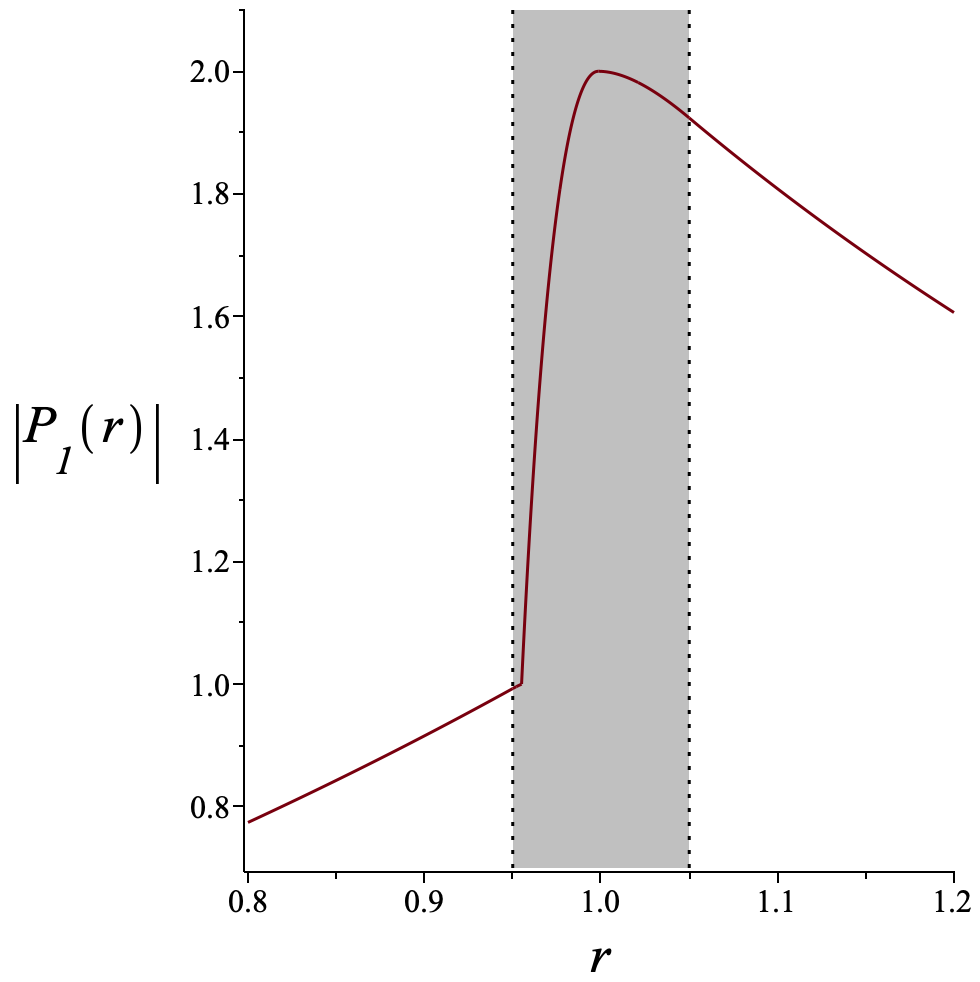}
   \caption{Modulus of the kink continuum eigenfunction $P_1$ for ratios $D_C/D_A=2$ (left) and $D_C/D_A=1/2$ (right), with $r_C=0.955R$, $k_z R=0.7$ and $l/R=0.1$. The values of $r$ are normalized to $R$. The linear transition profile defined by Eqs. \eqref{v_C^2Linear}-\eqref{v_s^2hat} is taken in the inhomogeneous layer, which is represented in gray.}
              \label{DCDA}%
    \end{figure}

As a comparison with the setup in Fig. \ref{eigenfM1}, we show the continuum eigenfunction of $P_1$ with a ratio $D_C/D_A=2$ and $D_C/D_A=1/2$ in Fig \ref{DCDA}. We see that the value of the ratio has an influence on the shape of the eigenfunction. It is not clear if all ratios are physically acceptable, or if an additional constraint fixes the ratio to one possible value.

\section{Conclusion}

In this paper, we investigated the slow surface mode in a straight cylinder with a circular base with an inhomogeneous layer included at the boundary through the eigenvalue problem. We extended the Frobenius series method used by \citet{SolerEtAl2013} for a different mode and in a different model, in order to adapt it to the situation where multiple series expansions are needed to cover the inhomogeneous layer. We were then able to find a dispersion relation for the eigenmodes of the cylinder. We first took the thin boundary limit of this relation, and recovered the approximative dispersion relation in the thin boundary approximation which was found by \citet{YuEtAl2017b}. Next, we investigated the full dispersion relation for a boundary layer of arbitrary width.

The inclusion of a finite inhomogeneous layer gives rise to the Alfv\'en and cusp continua in the frequency spectrum of the eigenmodes. A discrete mode having a frequency within one of these continua will then couple to a local continuum mode and become a damped global oscillation called a quasimode. Basing ourselves on the quasimodes studies through the Laplace transform of \citet{Sedlacek1971} and \citet{Goedbloed&Poedts2004} in order to find the quasimode from our dispersion function, we were able to find the complex frequency of the quasimode which corresponds to the slow surface mode of the discontinuous boundary case with its frequency in the cusp continuum.

We then proceeded to discuss an example case for the sausage mode, in which we took relatively simple transition profiles for the background MHD quantities in order for the analytical calculations to remain feasible. Since in resistive MHD the quasimode becomes an eigenmode, we compared our analytical results for the slow sausage mode to the numerical results obtained by \citet{ChenEtAl2018} from resistive MHD computations under the same model and the same background conditions (although with different transition profiles). Our findings were in line with the results and conclusions of that paper. In particular, the profile of the damping time-to-period ratio for our quasimode, important for characterizing the damping of a mode by resonant absorption, was found to be similar both in shape and in scale to their resistive slow sausage eigenmode. The small difference in scale is explained by the fact that resistivity is an additional damping mechanism with respect to ideal MHD. We also discussed the quasimode perturbations, and found that the jump in $P_1$ at the resonant position remains small, unlike in the case of the kink quasimode resonantly absorbed in the Alfv\'en continuum in a cold plasma as discussed by \citet{SolerEtAl2013}.

The case of nonaxisymmetric quasimodes was not extensively discussed in this paper. As the Alfv\'en and cusp continua overlap in the photospheric background conditions of interest in the present work, these modes will couple to continuum modes from both continua at the same time. The effect this has on the frequency of the quasimode not yet being well understood, we only briefly mentioned that case without delving into the details. However, this could be a subject for further study.

We also used the series method to plot the perturbation profiles of the sausage and kink ideal continuum modes corresponding to the resistive eigenmodes of \citet{GoossensEtAl2021}. We found that the continuum eigenfunctions match the resistive eigenfunctions relatively well, and suggest that the difference might be due to the inclusion of resistivity. In the case of the kink continuum modes, the simultaneous presence of both the cusp and Alfv\'en resonances leads to an apparent additional degree of freedom which influences the shape of the eigenfunctions. Further study on this subject is needed to determine whether or not an unknown factor removes this apparent degree of freedom and fixes the shape of the continuum eigenfunctions.

\begin{acknowledgements}
M.G. was supported by the C1 Grant TRACEspace of Internal Funds KU Leuven (number C14/19/089). TVD was supported by the European Research Council (ERC) under the European Union’s Horizon 2020 research and innovation programme (grant agreement No 724326) and the C1 grant TRACEspace of Internal Funds KU Leuven. This publication is part of the R+D+i project PID2020-112791GB-I00, financed by MCIN/AEI/10.13039/501100011033.
\end{acknowledgements}

%
%

\bibliographystyle{aa}
\bibliography{biblio}

\appendix

\section{Recursion formulas} \label{Recursions}

In this appendix, we derive the recursion formulas for the coefficients in front of the series from Section \ref{SolInh}. The coefficients are obtained by introducing the solutions in the form of either Frobenius series or power series into the ODE. 

We recall that, for the Frobenius solutions around the singularities of the ordinary differential equation (ODE) \eqref{eqP1} formed by the cusp and Alfv\'en resonant positions $r_C$ and $r_A$, the transition profiles of the squared cusp speed $v_C^2$ and the squared Alfv\'en speed $v_A^2$ in the inhomogeneous layer have been assumed to be strictly monotonic functions of $r$. Under this assumption, the indicial equations for the Frobenius series $\sum_{k=0}^{\infty} p_k (r-r_i)^{k+s}$ are $s(s-1)$ for $r_i=r_C$, and $s(s-2)$ for $r_i = r_A$ \citep{SakuraiEtAl1991}.

\subsection{Recursion formulas for the coefficients in $P_{1,1,C}$ and $P_{1,2,C}$}

For the Frobenius solution around the cusp resonant position $r_C$, we first rewrite the ODE \eqref{eqP1} as follows, by changing from the variable $r$ to the variable $\z=r-r_C$:

\begin{equation}
\dod[2]{P_1(\z)}{\z} + \d\frac{S(\z)}{\z} \dod{P_1(\z)}{\z} + \frac{Q(\z)}{\z^2} P_1(\z) = 0 \text{,}
\end{equation}
where $S$ and $Q$ are both analytic at $\z=0$ and are defined by

\begin{align}
S(\z) &=  \z \l\{ \d\frac{1}{\z + r_C} - \frac{\tod{}{\z} \l[ \rho_0 \l( \o^2 - \o_A^2 \r) \r]}{\rho_0 \l( \o^2 - \o_A^2 \r)} \r\} \text{,}\\
Q(\z) &= \z^2 \l\{ \d\frac{\l( \o_A^2 - \o^2 \r) \l( \o_s^2 - \o^2 \r)}{\l( \o_C^2 - \o^2 \r) \l( v_A^2 + v_s^2 \r)} - \frac{n^2}{\l(\z + r_C \r)^2} \r\} \text{.}
\end{align}
The indicial equation having roots $1$ and $0$ in this case, the first basic independent Frobenius solution around $r_C$ is given by $P_{1,1,C}(\z) = \sum_{k=0}^{\infty} \alpha_k \z^{k+1}$ and is called the small solution. Its coefficients $\alpha_k$ are determined by the following recursion relation:

\begin{align}
&\alpha_0 = \text{free} \\
&\alpha_k (k+1) k + \d\sum_{j=0}^k S_j \alpha_{k-j} (k-j+1) \notag \\
&+ \sum_{j=0}^k Q_j \alpha_{k-j} = 0 & \text{ for k $\geq$ 1,}
\end{align}
where $\a_0$ is free because we have one degree of freedom in each of the two independent solutions, whereas $S_j$ and $Q_j$ are the coefficients in front of $\z^j$ in the series expansions of respectively $S$ and $Q$ around $\z=0$. We note that $S_0 = S(0) = 0$ and $Q_0 = Q(0) = 0$, the latter following from the strict monotonicity of the transition profile $v_C^2$ in the inhomogeneous layer.

The second basic independent Frobenius solution around $r_C$ is given by $P_{1,2,C}(\z) = \sum_{k=0}^{\infty} \sigma_k \z^{k} + \mathcal{C}_C P_{1,1,C}(\z) \ln(\z)$ and is called the large solution. Its coefficients $\sigma_k$ and $\mathcal{C}_C$ are given by the recursion relation

\begin{align}
&\s_0 = \text{free} \\
&\mathcal{C}_C = -\d\frac{\s_0 Q_1}{\a_0} \text{,} \label{C_Cform}\\
&\s_1 = 0 \text{,} \\
&\s_{k} k(k-1) + \mathcal{C}_C (2k-1) \a_{k-1} \notag \\
&+ \d\sum_{j=0}^{k-1} \l[ \s_{j+1}(j+1) + \mathcal{C}_C \a_j \r] S_{k-j-1} \notag\\
& + \sum_{j=0}^{k} Q_{k-j} \s_{j} = 0 & \text{ for k $\geq$ 2,}
\end{align}
where $\s_0$ is free because we have one degree of freedom in each Frobenius series, and $\s_1$ can be assumed to be $0$ because otherwise $P_{1,2,C}(\z) - \frac{\s_1}{\a_0} P_{1,1,C}(\z)$ yields a new independent large Frobenius solution with the coefficient at the place of $\s_1$ equal to $0$.

\subsection{Recursion formulas for the coefficients in $P_{1,1,A}$ and $P_{1,2,A}$}

For the Frobenius solution around the Alfv\'en resonant position $r_A$, we first rewrite the ODE \eqref{eqP1} as follows, by changing from the variable $r$ to the variable $\ch=r-r_A$:

\begin{equation}
\dod[2]{P_1(\ch)}{\ch} + \d\frac{T(\ch)}{\ch} \dod{P_1(\ch)}{\ch} + \frac{U(\ch)}{\ch^2} P_1(\ch) = 0 \text{,}
\end{equation}
where $T$ and $U$ are analytic at $\ch=0$ and are defined by

\begin{align}
T(\ch) &=  \ch \l\{ \d\frac{1}{\ch + r_A} - \frac{\tod{}{\ch} \l[ \rho_0 \l( \o^2 - \o_A^2 \r) \r]}{\rho_0 \l( \o^2 - \o_A^2 \r)} \r\} \text{,}\\
U(\ch) &= \ch^2 \l\{ \d\frac{\l( \o_A^2 - \o^2 \r) \l( \o_s^2 - \o^2 \r)}{\l( \o_C^2 - \o^2 \r) \l( v_A^2 + v_s^2 \r)} - \frac{n^2}{\l(\ch + r_A \r)^2} \r\} \text{.}
\end{align}
The indicial equation having roots $2$ and $0$ in this case, the first basic independent Frobenius solution around $r_A$ is given by $P_{1,1,A}(\ch) = \sum_{k=0}^{\infty} \beta_k \ch^{k+2}$ and is called the small solution. Its coefficients $\beta_k$ are determined by the following recursion relation:

\begin{align}
&\beta_0 = \text{free,} \\
&\beta_k (k+2) (k+1) + \d\sum_{j=0}^k T_j \beta_{k-j} (k-j+2) \notag \\
&+ \sum_{j=0}^k U_j \beta_{k-j} = 0 & \text{ for k $\geq$ 1,}
\end{align}
where $\b_0$ is free because we have one degree of freedom in each Frobenius series, whereas $T_j$ and $U_j$ are the coefficients in front of $\ch^j$ in the series expansions of respectively $T$ and $U$ around $\ch=0$. We note that $U_0 = U(0) = 0$, $U_1 = \od{U}{\ch}\Bigr\rvert_{\ch=0} = 0$, and $T_0 = T(0) = -1$, the latter following from the strict monotonicity of the transition profile $v_A^2$ in the inhomogeneous layer.

The second basic independent Frobenius solution around $r_A$ is given by $P_{1,2,A}(\ch) = \sum_{k=0}^{\infty} \t_k \ch^{k} + \mathcal{C}_A P_{1,1,A}(\ch) \ln(\ch)$ and is called the large solution. Its coefficients $\t_k$ and $\mathcal{C}_A$ are given by the recursion relation

\begin{align}
&\t_0 = \text{free,} \\
&\mathcal{C}_A = -\d\frac{\t_0 U_2}{2 \b_0} \text{,}\\
&\t_1 = 0 \text{,} \\
&\t_2 = 0 \text{,} \\
&\t_{k} k(k-1) + \mathcal{C}_A (2k-1) \b_{k-2} \notag \\
&+ \d\sum_{j=0}^{k-1} \t_{j+1}(j+1) T_{k-j-1} + \mathcal{C}_A \d\sum_{j=0}^{k-2} \b_j T_{k-j-2} \notag\\
&+ \sum_{j=0}^{k} U_{k-j} \t_{j} = 0 & \text{ for k $\geq$ 3,}
\end{align}
where $\t_0$ is free because we have one degree of freedom in each Frobenius series, $\t_1=0$ because both $U_0=0$ and $U_1=0$, and $\t_2$ can be assumed to be $0$ because otherwise $P_{1,2,A}(\ch) - \frac{\t_2}{\b_0} P_{1,1,A}(\ch)$ yields a new independent large Frobenius solution with the coefficient at the place of $\t_2$ equal to $0$.

\subsection{Recursion formulas for the coefficients in $P_{1,1,r_0}$ and $P_{1,2,r_0}$}

For the solutions in the form of a power series around a regular point $r_0$, we write the ODE \eqref{eqP1} as

\begin{equation}
\dod[2]{P_1(r)}{r} + X(r) \dod{P_1(r)}{r} + Y(r) P_1(r) = 0 \text{,}
\end{equation}
where $X$ and $Y$ are defined by

\begin{align}
& X(r) = \frac{1}{r} - \frac{\tod{}{r} \l[ \rho_0 \l( \o^2 - \o_A^2 \r) \r]}{\rho_0 \l( \o^2 - \o_A^2 \r)} \text{,}\\
& Y(r) = \d\frac{\l( \o_A^2 - \o^2 \r) \l( \o_s^2 - \o^2 \r)}{\l( \o_C^2 - \o^2 \r) \l( v_A^2 + v_s^2 \r)} - \frac{n^2}{r^2} \text{.}
\end{align}
A basic independent solution is given by $\sum_{k=0}^{\infty} p_k (r-r_0)^{k}$, whose coefficients $p_k$ are determined by the following recursion relation:

\begin{align}
&p_0 = \text{free,} \\
&p_1 = \text{free,} \\
&p_{k} k (k-1) + \d\sum_{j=0}^{k-2} X_j p_{k-j-1} (k-j-1) \notag\\
& + \sum_{j=0}^{k-2} Y_j p_{k-j-2} = 0 & \text{ for k $\geq$ 2,}
\end{align}
where $p_0$ and $p_1$ are free because we have two degrees of freedom for a solution of a second-order ODE, whereas $X_j$ and $Y_j$ are the coefficients in front of $(r-r_0)^j$ in the series expansions of respectively $X$ and $Y$ around $r=r_0$. Two basic independent solutions are then obtained by choosing two different couples $(p_0, p_1)$ which are not multiples of one another.

\section{Deriving the expression for $\mathcal{C}_C$} \label{AppC_C}

In this appendix we derive expression \eqref{C_C} for $\mathcal{C}_C$. From Eq. \eqref{C_Cform} in Appendix \ref{Recursions}, we have that

\begin{equation} \label{C_CInApp}
\mathcal{C}_C = -\d\frac{\s_0 Q_1}{\a_0} \text{,}
\end{equation}
where we refer to the previous appendix for the definitions of $\a_0$, $\s_0$, and $Q_1$. Since $Q_1 = \tod{Q}{\z}\Bigr\rvert_{\z=0}$ (where we again refer to Appendix \ref{Recursions} for the definitions of $\z$ and $Q$), we find that

\begin{align}
Q_1 &= \dod{Q}{\z}\biggr\rvert_{\z=0} \\
&= \d\lim_{\z \to 0} \l\{ 2 \z F + \z^2 \dod{F}{\z'}\biggr\rvert_{\z'=\z} \r\} \label{limit}
\end{align}
with

\begin{equation}
F = \d\frac{\l( \o_A^2 - \o^2 \r) \l( \o_s^2 - \o^2 \r)}{\l( \o_C^2 - \o^2 \r) \l( v_A^2 + v_s^2 \r)} - \frac{n^2}{\l(\z + r_C \r)^2} \text{.}
\end{equation}

In order to proceed, we recall that $\o^2 = \o_C^2(\z=0)$ and make a Taylor expansion of $\o_C^2$ around $\z=0$ to find that

\begin{align}
&\o_C^2 - \o^2 = \dod{\o_C^2}{\z}\Biggr\rvert_{\z=0} \z + O(\z^2) \text{,} \\
&\l(\o_C^2 - \o^2 \r)^2 = \l( \dod{\o_C^2}{\z}\Biggr\rvert_{\z=0} \r)^2 \z^2 + O(\z^3) \text{,} \\
&\dod{}{\z} \l(\o_C^2 - \o^2 \r) = \dod{\o_C^2}{\z}\Biggr\rvert_{\z=0} + O(\z) \text{,}
\end{align}
as $\z \to 0$. We note that, since the transition profile of $v_C^2$ has been assumed to be strictly monotonic, $\tod{\o_C^2}{\z}\Bigr\rvert_{\z=0}$ will be different from $0$. We can then write the two terms in the limit on the right-hand side of Eq. \eqref{limit} as

\begin{equation} \label{term1}
2 \z F = \d\frac{2 \z \l(\o_A^2 - \o^2 \r) \l(\o_s^2 - \o^2 \r) }{\l( v_A^2 + v_s^2 \r) \l( \tod{\o_C^2}{\z'}\Bigr\rvert_{\z'=\z} \z + O(\z^2) \r)} - \frac{2 \z n^2}{\l( \z + r_C \r)^2}
\end{equation}
and

\begin{align}
&\z^2 \dod{F}{\z'}\biggr\rvert_{\z'=\z} =\notag\\
&\frac{\z^2 \l[\tod{F_1}{\z'}\Bigr\rvert_{\z'=\z} \l( v_A^2 + v_s^2 \r) \l(\o_C^2 - \o^2 \r) -F_1 F_2 \r]}{\l( v_A^2 + v_s^2 \r)^2 \l[ \l(\tod{\o_C^2}{\z'}\Bigr\rvert_{\z'=\z} \r)^2 \z^2 + O(\z^3) \r]} + \frac{2 \z^2 n^2}{\l( \z + r_C \r)^3} \text{,} \label{term2}
\end{align}
with

\begin{align}
F_1(\z) &= \l(\o_A^2 - \o^2 \r) \l(\o_s^2 - \o^2 \r) \text{,} \\
F_2(\z) &= \dod{}{\z}\l[\l(\o_C^2 - \o^2 \r) \l( v_A^2 + v_s^2 \r)\r] \\
&= \l( v_A^2 + v_s^2 \r) \l( \tod{\o_C^2}{\z'}\Bigr\rvert_{\z'=\z} + O(\z) \r) + \l(\o_C^2 - \o^2 \r) \dod{}{\z} \l( v_A^2 + v_s^2 \r) \text{.}
\end{align}

Hence, from Eqs. \eqref{limit}, \eqref{term1} and \eqref{term2}, we find that
\begin{align}
Q_1 = &\d\frac{2 \l(\o_A^2(0) - \o^2 \r) \l(\o_s^2(0) - \o^2 \r)}{\l( v_A^2(0) + v_s^2(0) \r) \tod{\o_C^2}{\z}\Bigr\rvert_{\z=0}} \notag\\
&- \frac{\l(\o_A^2(0) - \o^2 \r) \l(\o_s^2(0) - \o^2 \r) \l( v_A^2(0) + v_s^2(0) \r) \tod{\o_C^2}{\z}\Bigr\rvert_{\z=0}}{\l( v_A^2(0) + v_s^2(0) \r)^2 \l(\tod{\o_C^2}{\z}\Bigr\rvert_{\z=0}\r)^2} \\
= &\d\frac{\l(\o_A^2(0) - \o^2 \r) \l(\o_s^2(0) - \o^2 \r)}{\l( v_A^2(0) + v_s^2(0) \r) \tod{\o_C^2}{\z}\Bigr\rvert_{\z=0}}\\
= & \d\frac{k_z^4 v_C^4(0)}{\l(v_A^2(0) + v_s^2(0) \r) \tod{\o_C^2}{\z}\Bigr\rvert_{\z=0}}\text{,} \label{Q_1Final}
\end{align}
where the last equality holds because $\o^2 = \o_C^2(0) = k_z^2 v_C^2(0)$. 

Until now, the quantities were expressed as functions of $\z=r-r_C$. If we view the quantities as functions of the radial coordinate $r$, we obtain from Eqs. \eqref{C_CInApp} and \eqref{Q_1Final} that

\begin{equation}
\mathcal{C}_C = \d\frac{-\s_0 Q_1}{\a_0} = -\d\frac{\s_0 k_z^4 v_C^4(r_C)}{\a_0 \l(v_A^2(r_C) + v_s^2(r_C) \r) \tod{\o_C^2}{r}\Bigr\rvert_{r=r_C}} \text{,}
\end{equation}
which yields expression \eqref{C_C} when taking $\a_0 = \s_0 = 1$, as we did in Section \ref{FrobrC}.

\end{document}